\documentclass[journal,10pt]{IEEEtran}
\usepackage{amssymb}
\usepackage{amsmath}
\usepackage{cite}
\usepackage{url}
\usepackage{xcolor}
\usepackage{cite,graphicx,amsmath,amssymb}
\usepackage{subfigure}
\usepackage{citesort}
\usepackage{fancyhdr}
\usepackage{mdwmath}
\usepackage{mdwtab}
\usepackage{caption}
\usepackage{amsthm}
\usepackage{setspace}
\usepackage{booktabs}
\usepackage{algorithm}
\usepackage{algorithmic}
\usepackage{graphicx}

\newtheorem{remark}{Remark}
\newtheorem{theorem}{Theorem}

\newtheorem{lemma}{Lemma}

\newtheorem{corollary}{Corollary}

\makeatletter
\def\ScaleIfNeeded{%
\ifdim\Gin@nat@width>\linewidth \linewidth \else \Gin@nat@width
\fi } \makeatother

\begin{document}

\title{D2D-Enabled Mobile User Edge Caching: A Multi-Winner Auction Approach}


\author{

Tiankui~Zhang,~\IEEEmembership{Senior Member,~IEEE,}
        Xinyuan~Fang,
        Yuanwei~Liu,~\IEEEmembership{Senior Member,~IEEE,}
        Geoffrey~Ye~Li,~\IEEEmembership{Fellow,~IEEE,}
        and Wenjun~Xu,~\IEEEmembership{Senior Member,~IEEE}

\thanks{
This work was supported by National Natural Science Foundation of China under Grants 61971060 and 61502046.
Part of this work have been presented at the IEEE International Conference on Communication(ICC), Shanghai, China, May, 2019\cite{Fang2018}. (Co-corresponding authors: Tiankui Zhang and Wenjun Xu.)}
\thanks{Tiankui Zhang, Xinyuan~Fang and Wenjun~Xu are with
Beijing University of Posts and Telecommunications, Beijing, China (e-mail: \{zhangtiankui, fangxinyuan, wjxu\}@bupt.edu.cn).}
\thanks{Yuanwei Liu is with Queen Mary University of London, London, U.K. (e-mail:  yuanwei.liu@qmul.ac.uk).}
\thanks{Geoffrey Ye Li is with Georgia Institute of Technology, Atlanta, U.S. (e-mail: liye@ece.gatech.edu).}
\thanks{
Copyright (c) 2015 IEEE. Personal use of this material is permitted. However, permission to use this material for any other purposes must be obtained from the IEEE by sending a request to pubs-permissions@ieee.org.}
}
\maketitle
\begin{abstract}
In device-to-device (D2D)-enabled caching cellular networks, the user terminals (UTs) collaboratively store and share a large volume of popular contents from the base station (BS) for traffic offloading and delivery delay reduction. In this article, the multi-winner auction based caching placement in D2D-enabled caching cellular networks is investigated for UT edge caching incentive and content caching redundancy reduction. Firstly, a multi-winner once auction for UT edge caching is modeled which auctions multiple contents for multiple UTs. Then the optimization problem for content caching revenue maximization is formulated. Specifically, the  ``cache conflict" restriction relationship among UTs is used as one of the constraints in the problem to reduce the content caching redundancy in a UT movement scenario. The problem is solved by semidefinite programming (SDP) relaxation to obtain an approximate optimal caching placement. Moreover, the payment strategy of the auction is developed as a Nash bargaining game for personal profit fairness among the UTs who win the auction for content caching. Subsequently, a multi-winner once auction based caching (MOAC) placement algorithm is proposed. In addition, due to the high complexity of MOAC, we further propose a heuristic multi-winner repeated auction based caching placement (MRAC) algorithm, which can greatly reduce the complexity with only tiny performance loss. Simulation results show that the proposed algorithms can reduce the traffic load and average content access delay effectively compared with the existing caching placement algorithms.
\end{abstract}
\vspace{-0.3em}
\vspace{-1em}
\section{Introduction}
With the development of mobile network technologies and the popularization of mobile Internet applications, the mobile Internet data traffic and the content diversity have grown explosively recent years. It is predicted that data traffic in the global mobile cellular network will reach 2~ZB in 2021~\cite{whitepaper}, of which 71\% of mobile data traffic is used for content distribution. In order to meet the huge demands for content distribution in mobile cellular networks, the concept of edge caching has been proposed~\cite{Bastug2014Living,Golrezaei2013Femtocc}. Edge caching proactively stores hot contents with high popularity on base stations (BSs)~\cite{Bastug2014Living}, user terminals (UTs)~\cite{Golrezaei2013Femtocc} and unmanned aerial vehicles (UAVs)~\cite{ICInan,8254370}, thereby reducing content acquisition delay and backhaul link load~\cite{Wang2017A}. However, compared to the massive data contents, the cache spaces of nodes are limited. Therefore, the caching placement problem is the key to optimizing performance of edge caching.\par

The main idea of UT edge caching in cellular networks is to store the contents in some UTs by means of prefetching, and then share the contents among UTs by device-to-device (D2D) communications~\cite{Liu2016Cache}.  The combination of D2D communication and UT edge caching can effectively solve the problem of repeated content reuse during peak hours, improve the quality of user service, and reduce BS traffic in the hotspot area, which can be deployed in the campus of university as a typical application scenarios. The D2D communications can directly transmit data between UTs in the vicinity without passing through the BS by using intra-cell radio resource~\cite{6560489}. D2D communications in cellular networks can increase network throughput, reduce energy consumption, and improve spectrum utilization~\cite{Asadi2013A,6807946,7320989}. It has been shown that the throughput of caching cellular networks based on D2D communications increases linearly with the number of users~\cite{Ji2015Wireless}.
In fact, the caching placement is critical to caching performance in D2D-enabled caching since the cache space is limited compared to the large volume of  contents.
For the content diversity, one of the targets of the caching placement is to avoid redundancy and space waste.
In~\cite{Golrezaei2013Femtocc}, a wireless video storage distribution architecture using D2D in a small cell has been proposed to improve the video throughput. This architecture has used cache spaces instead of backhaul links in the applications with lots of content reuse and selected some UTs as fixed helper nodes. Based on the similar idea, the caching algorithm in~\cite{Kang2014Mobile} assigns a part of the UTs as  caching-server devices. In~\cite{Wang2015TASA}, a series of users willing to cache the same content are selected as the initial seeds according to their online influences and offline movements.
Some works select the helper nodes in the UTs according to their relationship in advance~\cite{Golrezaei2013Femtocc,Kang2014Mobile,Wang2015TASA}. However, the existing works only consider factors such as distance, interest similarity, or influence among users. The selected helper nodes are fixed and lack specificity for different contents.
In~\cite{Golrezaei2014Base}, the UTs in the same cluster cache different contents. But this approach cannot guarantee global optimality. In~\cite{tiankui}, a user preference aware caching deployment algorithm is proposed for D2D-enabled cellular networks with a constraint of one content can only saved in one UT in its coverage region. In~\cite{Li18}, the large-scale optimization and framework design of hierarchical edge caching are addressed which considers the analysis of social behavior and preference of mobile users, heterogeneous cache sizes, and the derived system topology in D2D-enabled cellular networks.
Besides, the above studies ignore the selfishness of UTs by assuming all UTs voluntarily participate in the content caching and sharing.
In practical applications, such as video sharing, news subscription, etc., content sharing among UTs through D2D communications needs more convincing reasons. Lack of incentives may prevent UTs to perform D2D communications.

Various incentive mechanisms have been designed for edge caching in cellular networks~\cite{LIU2018,Fang2018}.
Recently, there have been studies for caching placement based on auction theory or game theory in BS edge caching~\cite{Dai2012Collaborative,Hu2016Caching,Cao2017Mobile,8369389}. A  Vickrey-Clark-Groves (VCG) auction mechanism has been proposed in~\cite{Dai2012Collaborative} to maximize social welfare. This mechanism enables multiple wireless service providers (WPSs) to exchange cooperation buffers for bandwidth resources through the auction of server bandwidth. A multi-target auction has been designed in~\cite{Hu2016Caching}, where operators would bid for storage space for their own content. An auction algorithm is used in~\cite{Cao2017Mobile} to determine the amount of contents cached by the BSs. In~\cite{8369389}, stackelberg game for BS edge caching is designed for 5G networks with a large number of BSs and UTs.
For UE edge caching in D2D-enabled cellular networks, incentive mechanisms have been proposed for UTs participating in the caching placement process. Most of those research contributions  are  based on social relationships among UTs to enhance D2D communications~\cite{Chen2016Caching,Chen2015Exploiting,Zhu2017Social}.
The caching scheme in~\cite{Chen2016Caching} simultaneously enhances the participation of BSs and UTs by incentives, which utilizes game theory to maximize the revenue of UTs while minimize the cost of BSs. Some works have considered both content caching redundancy issues and user selfishness issues~\cite{Chen2015Exploiting,Zhu2017Social}. Social ties are used in~\cite{Chen2015Exploiting} to promote effective cooperation among UTs. Social-aware caching game has been proposed in~\cite{Zhu2017Social} to incentivize UTs to cache data for others.

Big data and artificial intelligence have been used for edge caching and D2D communication~\cite{wang18,wangwei17,coca18}. The potential of disseminating contents among mobile users assisted by the integration of big data techniques is discussed in~\cite{wang18}, but does not involve a specific caching placement strategy.
A distributed cache replacement strategy based on Q-learning at the BS has proposed in~\cite{wangwei17}. Joint optimization of computation, caching, and communication on the edge cloud has been proposed in~\cite{coca18}. However, both of them discuss the BS edge caching instead of UT edge caching we studies in the paper.

\vspace{-0.5em}
\subsection{Motivation and Contribution}
As mentioned above, the user selfishness problem and content caching redundancy should be considered in the caching placement design in D2D-enabled caching cellular networks.
However, some studies on caching placement overlook user selfishness. Even if user selfishness is considered in some works, the differences between various contents and the impact of user mobility are ignored.
In this paper, we consider both issues on UT edge caching in D2D-enabled caching  cellular networks.
We model the UT edge caching as a multi-winner once auction process, based on which we use the natural social efficiency and individual rationality characteristics of the auction mechanism to solve the problem of user selfishness.
By maximizing the content caching revenue of all the UTs in the cell, we develop a caching placement algorithm to cache different contents in different winner UT sets. Different from the prior helper nodes selection and clustering method, the UT sets are formed by the ``cache conflict" restriction relationship, which can reduce the content cache redundancy in the UT movement scenario. We design a content caching revenue function that considers content preferences, restriction relationships, and cache space costs of UTs. The proposed caching placement algorithm can effectively reduce the traffic load of the BS and average content access delay of the UTs, thereby improving the service experience of the UTs. The main contribution of this paper is summarized as follows,
\begin{itemize}
  \item We propose a multi-winner once auction of UT edge caching to incentivize the UTs participating content caching and sharing. Since there are multiple content chunks to be proactive cached in multiple UTs simultaneously at off-peak hours in practical application, the multiple content chunks regarded as products are auctioned to multiple UTs at one time in our proposed auction model.

   \item  We develop a caching placement optimization problem to maximize the content sharing revenue of all the UTs. Specifically, we utilize the  ``cache conflict" restriction relationship of UTs as one of the constraints in this problem to reduce the content cache redundancy in the UT movement scenario.
   We solve the formulated problem by semidefinite programming (SDP) relaxation and propose a corresponding multi-winner once auction based caching (MOAC) placement algorithm. Then we propose the pricing and payment strategy of the winner UT set for the auction. The payment is modeled as a Nash bargaining game, and the price paid to the content of each UT in the winner UT set is a Nash bargaining solution that can guarantee the fairness of personal profit of each UT.

  \item We propose a low-complexity multi-winner repeated auction caching (MRAC) placement algorithm to reduce the complexity of the MOAC. The MRAC sorts the content chunks according to their popularity, and then performs multi-winner auctions for different content chunks in sequence, which enable near-optimal caching placement, but greatly reduces complexity.

   \item We explore the impact of the encounter probability threshold on the performance of the proposed MOAC and MRAC algorithms, and analyzed the statistical characteristics of the encounter probability and the choice of its threshold. We demonstrate that the proposed algorithms are capable of achieving a near-optimal performance. Moreover, the MOAC and MRAC algorithms outperform the conventional caching placement schemes in terms of BS traffic offloading and content access delay under the appropriate threshold.
\end{itemize}
\vspace{-0.5em}
\subsection{Organization and Notations}
The rest of the paper is organized as follows. In Section II, D2D-enabled caching cellular networks considering  UT mobility is introduced. The MOAC placement algorithm is presented in Section III. The MORC placement algorithm is addressed in Section IV. Simulation results are shown in Section V and conclusions are finally drawn in Section VI. The main symbols and variables used in this paper are summarized in Table I.
\begin{table}[!htbp]
\vspace{-0.5em}
\small
  \begin{center}
      \caption{Main Symbol and Variable List}
   \begin{tabular}{|c|p{6.3cm}<{\centering}|}
      \hline
      \textbf{Parameter} & \textbf{Description}  \\       \hline
      $M$ & total number of content chunks in one macrocell \\ \hline
      $N$ & total number of UTs in one macrocell \\ \hline
      $v_n^{\left( m \right)}$ & the content caching and sharing revenue of UT $n$ for content chunk $m$ \\ \hline
      $b_n^{\left( m \right)}$ & the bid of UT $n$ for content chunk $m$ \\ \hline
      $p_n^{\left( m \right)}$ & the price that winner UT $n$ needs to pay for content chunk $m$ \\ \hline
      ${{\bf{p}}^{\left( m \right)}}$ & ${{\bf{p}}^{\left( m \right)}} = \left[ {p_1^{\left( m \right)},p_2^{\left( m \right)},...,p_N^{\left( m \right)}} \right]$ \\ \hline
      $x_n^{\left( m \right)}$ & $x_n^{\left( m \right)} = 1$ represents UT $n$ caching the content chunk $m$ \\ \hline
      ${{\bf{x}}^{\left( m \right)}}$ & ${{\bf{x}}^{\left( m \right)}} = \left[ {x_1^{\left( m \right)},x_2^{\left( m \right)},...,x_N^{\left( m \right)}} \right]$ \\ \hline
      ${e_{n,n'}}$ &  encounter probability between UT $n$ and $n'$ \\ \hline
      ${\bf{E}}$ & UT ``cache conflict" restriction matrix \\ \hline
      $\gamma$ & encounter probability threshold \\ \hline
      ${{\cal N}_n}$ & the neighboring UT set of UT $n$ \\ \hline
      $w_n^{\left( m \right)}$ & the content sharing profit of UT $n$ for content chunk $m$ \\ \hline
      $f_{n}^{\left( m \right)}$ & preference of UT $n$ for content chunk $m$ \\ \hline
      ${\bf{\chi }}$ & $\chi {\rm{ = }}\left[ {{{\bf{x}}^{\left( 1 \right)}},{{\bf{x}}^{\left( 2 \right)}},...,{{\bf{x}}^{\left( M \right)}}} \right]$ \\ \hline
      $\upsilon$ &  ${\bf{\upsilon }} = [{{\bf{v}}^{\left( 1 \right)}},{{\bf{v}}^{\left( 2 \right)}},...{{\bf{v}}^{\left( M \right)}}]$ \\ \hline
      ${{\cal N}^{\left( m \right)}}$ & the winner set for content chunk $m$ \\ \hline
      $\cal W$ & the winner sets include all content chunks \\ \hline
      ${F_m}$ & local popularity of content chunk $m$ \\ \hline
     \end{tabular}
     \vspace{-0.5em}
  \end{center}
 \vspace{-1em}
\end{table}

\vspace{-0.5em}
\section{System Model}
\vspace{-0.5em}
\subsection{Cellular Networks}
\begin{figure} [thp]
\centering
\includegraphics[width=0.98\linewidth]{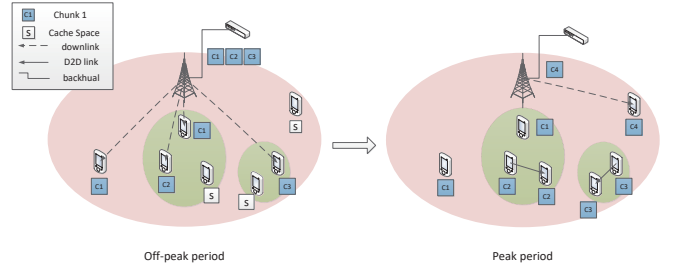}
 \caption{D2D-enabled caching cellular networks. }
 \label{system_model}
 \vspace{-0.5em}
\end{figure}

The D2D-enabled caching cellular networks considered in this paper are shown in Fig.~\ref{system_model}, there is one cell with one macro BS and a set of UTs,  ${\cal N}{\rm{ = \{ }}1,2,...,N{\rm{\} }}$. It is assumed that multiple UTs are randomly distributed within the coverage of the BS. The UTs can communicate with each other by  D2D communication links.
In this article, we assume that the content-oriented information management function, including content naming and caching placement/replacement decision is embedded in the macro BS in each cell, so we focus on the caching placement in one cell. In fact, the proposed algorithms is can be extended to multiple cells scenario directly if the management function is executed by a central baseband unit (BBU) in a cloud RAN (CRAN) architecture.

It is assumed that the cellular transmissions and D2D transmission are assigned non-overlapping orthogonal radio resource, which is referred to as orthogonal mode or an overlay mode.
The system bandwidth of downlink cellular transmissions is $B_C$ and the system bandwidth of downlink D2D transmission is $B_D$. We use the equal radio resource allocation for both the cellular UTs and the D2D UTs.

When UT $n'$ is communicating with UT $n$ for content delivery, the data rate from UT $n'$ to UT $n$ in time slot $t$ is
 \begin{equation}
{{r_{n,n'}}\left( t \right)}
 = {\frac{B_D}{N_D}}\log \left( {1 + \frac{{{g_{n',n}}\left( t \right)P_{n'}^{tx}}}{{{{I_D\left( t \right)}+\sigma ^2}}}} \right);
\end{equation}
when the BS is communicating with UT $n$ for content delivery, the data rate from its serving BS to UT $n$ in time slot $t$ is
 \begin{equation}
{r_{n,BS}}\left( t \right) = {\frac{B_C}{N_C}}\log \left( {1 + \frac{{{g_{BS,n}}\left( t \right)P_{BS}^{tx}}}{{{{I_C\left( t \right)}+\sigma ^2}}}} \right),
\end{equation}
where ${\sigma ^2}$ is additive white Gaussian noise power,
$N_C$ is the number of cellular UTs receiving data from BS,
$N_D$ is the number of D2D communication UT pairs,
$P_{n'}^{tx}$ and  $P_{BS}^{tx}$ is the maximum transmit power of UT $n'$ and the maximum transmit power of BS, ${g_{n',n}}\left( t \right)$ and ${g_{BS,n}}\left( t \right)$ are the channel gains between UT $n'$ and UT $n$ and between BS and UT $n$ in time slot $t$, $I_C\left( t \right)$ and $I_D\left( t \right)$ are co-channel interference levels from adjacent cells in time slot $t$  by cellular communications and D2D communications, respectively.
The inter-cell interference coordination by power allocation and channel scheduling in D2D cellular networks can improve the system performance, as shown in~\cite{ICIchannel,ICIpower}. Since we focus on the caching placement, the interference coordination is out of the scope of this paper.

It is noted that, the proposed caching placement algorithms can be used in a D2D communication system working in the non-orthogonal mode (underlay mode), which means that
the cellular transmissions and D2D transmission share the same radio resources. However, D2D communications in the non-orthogonal mode suffer strong interference among cellular UTs and D2D UTs, which can be solved by nonorthogonal multiple access (NOMA) techniques~\cite{liu1,liu2}.
\vspace{-0.5em}
\subsection{Caching Model}
In the mobile networks, we assume that each content file is divided into several unified chunks, which is the minimal unit of data to be transferred over the network. We set the size of each content chunk is $s$. It means that different content files with various data size can be divided into various chunks.
We assume that each cell can cache
a set of content chunks ${\cal M}{\rm{ = \{ }}1,2,...,M{\rm{\} }}$.
Since the UTs¡¯ caching spaces are limited, we assume that each UT can only store a fewer number of content chunks, the cache space of a UT is $s_0$.

In our caching model, a mobile user can be a content requester and a content provider, as shown in Fig.~\ref{system_model}.
In the caching placement process, we assume that the cached files are popular audio and video contents whose popularity changes are much slower than the UTs movement, i.e. the content popularity is invariant during the UTs movement.
If UT $n$ caches the chunk $m$, $x_n^{\left( m \right)} = 1$; $x_n^{\left( m \right)} = 0$ otherwise. If there exists a complete copy of content chunk $m$ in its own cache, the request is fulfilled with no delay and without the need to establish a communication link. Otherwise, the UT broadcasts a request message for chunk $m$ to the neighbor UTs. If the UT can find the requested chunk from a UT's cache space within its D2D transmission range, then it will establish a D2D communication link and obtain the chunk. If the UT cannot find the requested chunk neither in its own cache nor its proximity UTs, it will access to BS to acquire this chunk.  Since one file is divided into multiple chunks, one UT can acquire the requested file from multiple adjacent UTs and BS.

\subsection{Mobility Model}
We study the caching placement problem when UTs are moving within a fixed area, such as a university campus, a residential area, and so on. In order to reduce the impact of UTs mobility on system performance, the caching placement is designed based on the encounter probabilities between UTs.

We use the clustered random model~\cite{4757211} to model the mobility of UTs. We assume that each UT has a home-point, and most of its activities are near this point. The home-point can be the UTs' dormitory or workplace. We assume that a UT moves around its home-point according to a general ergodic process. This UT movement process will result into a rotationally-invariant spatial distribution $\phi \left( d \right)$ which decays with the distance $d$ between the UT and its home-point. For a given two points ${{X_1}}$ and ${{X_2}}$ within the coverage of a BS of radius $R$, the distance $d$ between them is defined as,
\begin{equation}
\small
d\left( {{X_1},{X_2}} \right) = \mathop {\min }\limits_{u.v \in \left\{ { - R,0,R} \right\}} \sqrt {{{\left( {{x_1} + u - {x_2}} \right)}^2} + {{\left( {{y_1} + v - {y_2}} \right)}^2}}.
\end{equation}

In clustered random model, we define function $s\left( d \right) = \min \left( {1,{d^{ - \delta }}} \right)$ to avoid convergence problems in proximity of the home-point, and normalize it so as to obtain a proper probability density function over the network area ${{O_N}}$: $\phi \left( d \right) = \frac{{s\left( d \right)}}{{\int {\int_{{O_N}} {s\left( d \right)} } }}$. The normalization constant $G\left( n \right) = \int {\int_{{O_N}} {s\left( d \right)} } $ can be approximated in polar coordinates
\begin{equation}
\small
G\left( n \right) = \Theta \left( {\int_0^{2\pi } {{\rm{d}}\theta \int_0^R {\min \left( {1,{\rho ^{ - \delta }}} \right)\rho {\kern 1pt} {\kern 1pt} {\rm{d}}\rho } } } \right).\footnote{Given two functions $f\left( n \right) \ge 0$ and $g\left( n \right) \ge 0$: $f\left( n \right) = O\left( {g\left( n \right)} \right)$ means $\lim {\sup _{n \to \infty }}f\left( n \right)/g\left( n \right) = c < \infty $; $f\left( n \right) = \Theta \left( {g\left( n \right)} \right)$ means $f\left( n \right) = O\left( {g\left( n \right)} \right)$ and $g\left( n \right) = O\left( {f\left( n \right)} \right)$.}
\end{equation}
The average distance between the UT and its home-point can be approximated by
\begin{equation}
\small
\mathbb{E}\left[ d \right] = \Theta \left( {\frac{1}{{G\left( n \right)}}\int_0^{2\pi } {{\rm{d}}\theta \int_0^R {\min \left( {1,{\rho ^{ - \delta }}} \right)\rho {{\kern 1pt} ^2}{\kern 1pt} {\rm{d}}\rho } } } \right).
\end{equation}
Parameter $\delta$ accounts for the fact that an individual node does not visit uniformly the network area, but spends most of the time just in a limited portion of it. In clustered random model, We assume that there are several clusters with high density of home-points in the network area, such as school buildings and dormitory buildings on campus. Each cluster has a middle point that is uniformly located within ${{O_N}}$. The home-point of each UT is considered to belong to one of the clusters. Considering the reality, the home-points of some UTs may coincide. The home-points of the same cluster are then uniformly and independently placed within a disk of radius $R'~\left(R'<R\right)$ centered at the cluster middle point.

Based on such mobility model, we denote ${e_{n,n'}}$ as the probability of the encounter between UT $n$ and $n'$  in a given area, which is a known statistical result in a certain duration, such as $T$ time slots. In time slot $t$, UT $n$ encounters UT $n'$ means UT $n$ is in the D2D communication area of UT $n'$, that is, the received signal power of the D2D communication between $n$ and $n'$ is greater than a minimum received signal power level $K$. We denote ${c_{n,n'}}\left( t \right) = 1$ if UT $n$ encounters UT $n'$ in time slot $t$; otherwise ${c_{n,n'}}\left( t \right) = 0$. Then the encounter probability of UT $n$ and $n'$ during $T$ time slots is,
 \begin{align}\label{encounter}
{e_{n,n'}} = \frac{1}{T}\sum\limits_{t = 1}^T {{c_{n,n'}}\left( t \right)},
\end{align}
where $0 \le {e_{n,n'}} \le 1$. The larger ${e_{n,n'}}$ is, the higher the probability that $n$ and $n'$ can share contents through D2D communication.

 \vspace{-0.5em}
\section{Multi-Winner Once Auction Based Caching Placement}
In this section, the multi-winner once auction of UT edge caching is modeled first. Based on this auction model, the caching placement problem is formulated and its approximate optimal solution is obtained, then the pricing and payment strategy of this auction is given.
\vspace{-0.5em}
\subsection{Multi-Winner Auction Model}
In order to motivate UTs for content caching and sharing, the caching placement at UTs is modeled as an auction process. In our content auction model, the buyers (bidders) are the UTs, the seller is the BS,  the auction products are the content chunks.
There are three basic principles need to be met when designing an auction mechanism:
\begin{itemize}
 \item social efficiency: the social welfare of all participants is maximized;

 \item individual rationality: the revenue of each participant is larger than or equal to zero;

 \item truthfulness: the auction mechanism should ensure that the buyer¡¯s bid for the auction products is equal to the products' actual value.
\end{itemize}

In this paper, we consider the revenue of all the UTs  as social welfare, so the social efficiency can be ensured by maximizing the sum revenue of UTs.
We assume that only when the UT's revenue for caching a content chunk is greater than 0, this UT will participate in the auction of this content chunk. This assumption is feasible in practical scenario and is in line with individual rationality of each~UT.

There are many common auction models that can guarantee truthfulness, including VCG auction models and second-price sealed bid auction models. The existing auction research on spectrum access shows that VCG auctions are not conducive to cooperation and may result in lower returns~\cite{Wu2008A}. Therefore, we apply the second-price sealed bid auction in this paper.

The second-price sealed bid auction, also known as the Vickery auction, has three characteristics as follows:
\begin{itemize}
 \item the bidders cannot know (at least when bidding) the bids of other bidders;

 \item the bidders only bid once;

 \item the highest bidder wins, and the winner pays the second highest bid.
\end{itemize}

The advantage of the second-price sealed bid auction is that it can guarantee the truthfulness of bidders' bids because the winner's payment is not determined by themselves but by the second highest loser. The level of the bid only determines the winning or losing but does not determine the specific price. From~\cite{Krishna2010Auction}, we can see that only bidding with real values is the best strategy, which can maximally ensure the bidder¡¯s profit.

Considering individual rationality and truthfulness, the basic auction model we use is the second-price sealed bid auction. However, the typical mode of the traditional second-price sealed bid auction is the single winner auction. In our caching model, we assume that the same content chunk can be cached in multiple UTs. In doing so, we propose a multi-winner once auction (MOA) model based on second-price sealed bid auction, which achieves multiple winners by once auction for different UTs caching the same content chunk.

In the proposed MOA model for content caching and sharing, the revenue of UT $n$  for caching content chunk $m$ is $v_n^{\left( m \right)}$, the bid of UT $n$ for content chunk $m$ is $b_n^{\left( m \right)}$, and the winner UT $n$ needs to pay $p_n^{\left( m \right)}$ for the content chunk $m$.
As mentioned above, we consider that $b_n^{\left( m \right)} = v_n^{\left( m \right)}$ because the second-price sealed bid auction model can guarantee the truthfulness of the bidders.
The UTs win the same content form a winner UT set and each set can be viewed as one bidder. By doing so, the process of a multi-winner auction can be converted to a single-winner auction.
There is a predefined ``cache conflict" restriction relationship of UTs participating in the auction process. The winner is the set of UTs with the highest total bid without ``cache conflicts".
There are $M$ winner UT sets for $M$ content chunks, which means the $M$ content chunks are cached in the $M$ winner UT sets, then the caching placement vectors are denoted as ${{\bf{x}}^{{{\left( 1 \right)}}}}{\bf{,~}}{{\bf{x}}^{{{\left( 2 \right)}}}}{\bf{,~}}...{\bf{,~}}{{\bf{x}}^{\left( M \right)}}$.  Correspondingly, there are $M$ payment price vectors ${{\bf{p}}^{{{\left( 1 \right)}}}}{\bf{,~}}{{\bf{p}}^{{{\left( 2 \right)}}}}{\bf{,~}}...{\bf{,~}}{{\bf{p}}^{\left( M \right)}}$.
For the caching placement, the main work in this proposed MOA determines two parts:

1) the set of winners by  the social welfare maximization of all participants to obtain the optimal caching placement.

2) the price that UTs in the winner set need to pay, respectively, and get the pricing payment strategy.
\vspace{-0.5em}
\subsection{Problem Formulation of Caching Placement}
The purpose of caching placement in D2D cellular networks is to reduce the BS traffic load in peak hours and reduce the content access delay of mobile users. In doing so, we should cache as many contents as possible and take full advantage of the D2D communications for content delivery.

First, we define the ``cache conflict" restriction relationship of UTs to improve the cache utility by avoiding caching same contents repeatedly on multiple proximal UTs. The defined ``cache conflict" restriction relationship of UTs is decided by the encounter probabilities of UTs since the mobile users may move frequently in a certain resident area or university campus, in the practical scenario.
When the encounter probability of two UTs is large, the content sharing can be achieved by D2D communication between these two UTs. In such case, caching the same content in these two UTs leads a waste of the cache space and diminishes the content diversity in the networks. So we define a ``cache conflict" restriction of UTs to avoid the waste of the cache space.

The ``cache conflict" restriction matrix is defined as ${\bf{E}} = \left\{ {{E_{n,n'}}} \right\}$, where ${E_{n,n'}} = 1$ when
the encounter probability of UT $n$ and UT $n'$, ${e_{n,n'}}$, is larger than a predefined encounter probability threshold $\gamma$, i.e.,
 ${e_{n,n'}} > \gamma $. Otherwise, ${E_{n,n'}} = 0$. The ``cache conflict" restriction relationship of UTs is, the UT $n$ and UT $n'$ cannot cache the same content when ${E_{n,n'}} = 1$, that is,
\begin{align}\label{cache_conflict}
x_n^{\left( m \right)} + x_{n'}^{\left( m \right)} \le 1,{\kern 1pt} {\kern 1pt} {\kern 1pt} {\kern 1pt} {\kern 1pt} {\kern 1pt} {\kern 1pt} {\kern 1pt} {\rm{if}}{\kern 1pt} {\kern 1pt} {\kern 1pt} {\kern 1pt} {\kern 1pt} {\kern 1pt} {\kern 1pt} {\kern 1pt} {E_{n,n'}} = 1.
\end{align}
UT $n'$ is regards as a neighbor of UT $n$ when ${E_{n,n'}}=1$. The neighboring UT set of UT $n$ is denoted as ${{\cal N}_n}$.
In practical scenarios, the predefined encounter probability threshold, $\gamma$, can be obtained from the statistical data of the encounter probability of UTs. We will verify the impact of $\gamma$ on the performance of caching placement by simulation in Section~V.B.
From the definition of the ``cache conflict", $\gamma$ determines the sparseness of matrix ${\bf{E}}$. When $\gamma=1$, ${\bf{E}}$ is a zero matrix, which means that there is no ``cache conflict" between the UTs. When $\gamma=0$, almost all elements in ${\bf{E}}$ are equal to 1, which means the ``cache conflict" condition between UTs is very strict, that is, for each content chunk, it can be cached by only one UT in the cell.

The revenue of UT $n$ for content chunk $m$ is defined as
\begin{align}\label{value}
v_n^{\left( m \right)} = w_n^{\left( m \right)} - \zeta  {c_n},
\end{align}
where $w_n^{\left( m \right)}$ denotes the content sharing profit of UT $n$ for content chunk $m$, ${c_n}$ denotes the occupied cache space by content chunk $m$ in UT $n$, and $\zeta $ denotes the cost of the unit cache space. The revenue vectors for different content chunks are defined as ${{\bf{v}}^{\left( 1 \right)}},~{{\bf{v}}^{\left( 2 \right)}},~...,~{{\bf{v}}^{\left( M \right)}}$.
The content sharing profit of UT $n$ by D2D transmission in its neighboring UT set  ${{\cal N}_n}$ is defined as
\begin{align}\label{w}
w_n^{\left( m \right)} = \vartheta s\sum\limits_{n' \in {{{\cal N}_n}}} {f_{n'}^{\left( m \right)}{{\bar r}_{n,n'}}{e_{n,n'}}},
\end{align}
where $f_{n'}^{\left( m \right)}$ denotes the preference of UT $n'$ for content chunk $m$, and ${\bar r_{n,n'}} = \frac{1}{T}\sum\limits_{t = 1}^T {{r_{n,n'}}\left( t \right)}$. ${\bar r_{n,n'}}$ denotes the average transmission rate between UT $n$ and $n'$, ${e_{n,n'}}$ denotes the encounter probability of UT $n$ and $n'$, and $\vartheta$ denotes the transmission cost of unit bit content.

\vspace{-0.5em}
\begin{remark}\label{remark1}
The larger revenue means that the contents have a larger opportunity to be delivered by D2D communications. Therefore maximizing total revenue of all UTs means selecting the set of UTs with the highest encounter probabilities and the highest average transmission rate for content caching and sharing, which can effectively utilize D2D communication to reduce the traffic load of BS and content access delay of UTs.
\end{remark}
\vspace{-0.5em}
According to the proposed MOA model, the caching placement vectors, ${{\bf{x}}^{{{\left( 1 \right)}}}}{\bf{,~}}{{\bf{x}}^{{{\left( 2 \right)}}}}{\bf{,~}}...{\bf{,~}}{{\bf{x}}^{\left( M \right)}}$, can be obtained via once auction by maximizing the social welfare of all the UTs, which is formulated as follows:
\begin{subequations}\label{optimal1}
\begin{align}
&{U^*} = \mathop {\max }\limits_{{{\bf{x}}^{{{\left( 1 \right)}}}}{\bf{,}}{{\bf{x}}^{{{\left( 2 \right)}}}}{\bf{,}}...{\bf{,}}{{\bf{x}}^{{{\left( M \right)}}}}} \sum\limits_{m = 1}^M {\sum\limits_{n = 1}^N {v_n^{\left( m \right)}x_n^{\left( m \right)}} } \\
\label{10b}
&{\rm{s}}{\rm{.t}}{\rm{.}}{\kern 1pt} {\kern 1pt} {\kern 1pt} {\kern 1pt}x_n^{\left( m \right)} + x_{n'}^{\left( m \right)} \le 1,{\kern 1pt} {\kern 1pt} {\kern 1pt} {\kern 1pt} \forall n,n'{\kern 1pt} {\kern 1pt} {\kern 1pt} {\kern 1pt} {\rm{if}}{\kern 1pt} {\kern 1pt} {\kern 1pt} {\kern 1pt} {E_{n,n'}} = 1,{\kern 1pt} {\kern 1pt} {\kern 1pt} {\kern 1pt} \forall m,\\
\label{10c}
&{\kern 1pt} {\kern 1pt} {\kern 1pt}{\kern 1pt}{\kern 1pt}{\kern 1pt}{\kern 1pt}{\kern 1pt}{\kern 1pt}{\kern 1pt}{\kern 1pt}{\kern 1pt} {\kern 1pt}{\kern 1pt}{\kern 1pt}{\kern 1pt}{\kern 1pt}\sum\limits_{m = 1}^M {x_n^{\left( m \right)}s \le {s_0},} {\kern 1pt}{\kern 1pt}{\kern 1pt}{\kern 1pt} \forall n,\\
\label{10d}
&{\kern 1pt}{\kern 1pt}{\kern 1pt}{\kern 1pt}{\kern 1pt}{\kern 1pt}{\kern 1pt}{\kern 1pt}{\kern 1pt} {\kern 1pt} {\kern 1pt} {\kern 1pt} {\kern 1pt} {\kern 1pt} {\kern 1pt}{\kern 1pt}{\kern 1pt}{\kern 1pt}{\kern 1pt}{\kern 1pt}{\kern 1pt}x_n^{\left( m \right)} = \left\{{\rm{0}}, {\rm{1}}
\right\}, {\kern 1pt}{\kern 1pt}{\kern 1pt}{\kern 1pt}
{n \in {\cal N}}, {\kern 1pt} {\kern 1pt} {\kern 1pt} {\kern 1pt} {m \in {\cal M}},
\end{align}
\end{subequations}
where~\eqref{10b} denotes the ``cache conflict" restrict relationship of UT $n$ and UT $n'$,
\eqref{10c} constrains that the cached content chunks in a UT cannot exceed its cache space $s_0$,
and ~\eqref{10d} constrains that $x_n^{\left( m \right)}$ is a binary variable.
By solving this problem to maximum total revenue of all UTs for content caching and sharing, we obtain the optimal caching placement ${{\bf{x}}^{{{\left( 1 \right)}}}}{\bf{,~}}{{\bf{x}}^{{{\left( 2 \right)}}}}{\bf{,~}}...{\bf{,~}}{{\bf{x}}^{{{\left( M \right)}}}}$ and winner sets for different content chunks ${{\cal N}^{\left( 1 \right)}},~{{\cal N}^{\left( 2 \right)}},~...,~{{\cal N}^{\left( M \right)}}$.

In generally, the cache space of a UT is limited, so it can only cache a few content chunks, as indicated in~\cite{Umrao2018}. In this paper, for simplicity but without loss of generality, it is assumed that one UT can cache one content chunk with a size of $s$. If a UT has cache space more than one chunk, it is equivalent to multiple UTs with a unified cache space size~$s$.

However, the problem~\eqref{optimal1} is a maximal weighted independent set (MWIS) problem in graph theory.
As the core issue in the NP-complete class of problems, its simplest case with all ${v_n^{\left( m \right)}}=1$ has been proved to be NP-complete in~\cite{Karp}. Except for some special cases, such as when the graph is perfect (the problems involved in this paper are not such special cases), the complexity of MWIS is in general very high. Therefore, we need to find an approximate optimal solution of problem~\eqref{optimal1}.

\subsection{Approximate Optimal Solution of Caching Placement}
In order to simplify problem~\eqref{optimal1}, we regard the optimization problem of $M$ content chunks and $N$ UTs as the optimization problem of one content chunk ($M=1$) and $MN$ UTs participating.
For this purpose, we introduce two auxiliary variables ${\bf{\chi }}{\rm{ = }}\left[ {{{\bf{x}}^{\left( 1 \right)}},~{{\bf{x}}^{\left( 2 \right)}},~...,~{{\bf{x}}^{\left( M \right)}}} \right]$, and ${\bf{\upsilon }} = [{{\bf{v}}^{\left( 1 \right)}},~{{\bf{v}}^{\left( 2 \right)}},~...,~{{\bf{v}}^{\left( M \right)}}]$, then the problem~\eqref{optimal1} can be expressed equivalently as follows:
\begin{subequations}\label{optimal2}
\begin{align}
&{U^*} = \mathop {\max }\limits_{\chi } \sum\limits_{n = 1}^{MN} {{\upsilon _n}{\chi _n},} \\
&{\rm{s}}{\rm{.t}}{\rm{.}}{\kern 1pt} {\kern 1pt} {\kern 1pt} {\kern 1pt} {\chi _n} + {\chi _{n'}} \le 1,\forall n,n'{\kern 1pt} {\kern 1pt} {\kern 1pt} {\rm{if}}{\kern 1pt} {\kern 1pt} {\kern 1pt} {\kern 1pt} {\kern 1pt} {\kern 1pt} {\Xi _{n,n'}} = 1,\\
&{\kern 1pt} {\kern 1pt} {\kern 1pt} {\kern 1pt} {\kern 1pt} {\kern 1pt} {\kern 1pt} {\kern 1pt} {\kern 1pt} {\kern 1pt} {\kern 1pt} {\kern 1pt} {\kern 1pt} {\kern 1pt} {\kern 1pt} {\kern 1pt} {\kern 1pt} {\kern 1pt} {\kern 1pt} {\kern 1pt}{\chi _n} = \left\{{\rm{0}}, {\rm{1}} \right\}, n = 1,~2,~...,~MN,
\end{align}
\end{subequations}
where
\begin{align}\label{optimal22}
\small
\Xi  \buildrel \Delta \over = {\left[ {\begin{array}{*{20}{c}}
{\bf{E}}&{\bf{I}}& \cdots &{\bf{I}}\\
{\bf{I}}&{\bf{E}}& \cdots &{\bf{I}}\\
 \vdots & \vdots & \ddots & \vdots \\
{\bf{I}}&{\bf{I}}& \cdots &{\bf{E}}
\end{array}} \right]_{MN \times MN}}
\end{align}
is the expanded ``cache conflict" matrix of the $MN$ UTs based on ${\bf{E}}$, and ${\bf{I}}$ denotes the identity matrix.

The following lemma, proved in Appendix A, can simplify the problem~\eqref{optimal2} further.
\begin{lemma}\label{lemma:optimal2}
\emph{Define ${\bf{\mu }} = {[\sqrt {{\upsilon _1}} ,~\sqrt {{\upsilon _2}} ,~...,~\sqrt {{\upsilon _{MN}}} ]^T}$, ${\bf{y}} = \left[ {c\sqrt {{\upsilon _1}} {\chi _1},~c\sqrt {{\upsilon _2}} {\chi _2},~...,~c\sqrt {{\upsilon _{MN}}} {\chi _{MN}}} \right]$, $c$ is the normalization constant. Then the optimization problem~\eqref{optimal2} can be equivalent to the following optimization problem,}
\begin{subequations}\label{optimal3}
\begin{align}
&{{\tilde U}^*} = \mathop {\max }\limits_{\bf{y}} {\left( {{{\bf{\mu }}^T}{\bf{y}}} \right)^2},\\
&{\rm{s}}{\rm{.t}}{\rm{.}}{\kern 1pt} {\kern 1pt} {\kern 1pt} {\kern 1pt} {\kern 1pt} {\kern 1pt} {\kern 1pt} {\kern 1pt} {y_n}{y_{n'}} = 0,\forall n,n'{\kern 1pt} {\kern 1pt} {\kern 1pt} {\kern 1pt} {\rm{if}}{\kern 1pt} {\kern 1pt} {\kern 1pt} {\kern 1pt} {E_{n,n'}} = 1,\\
&{\kern 1pt} {\kern 1pt} {\kern 1pt} {\kern 1pt} {\kern 1pt} {\kern 1pt} {\kern 1pt} {\kern 1pt} {\kern 1pt} {\kern 1pt} {\kern 1pt} {\kern 1pt} {\kern 1pt} {\kern 1pt} {\kern 1pt} {\kern 1pt} {\kern 1pt} {\kern 1pt} {\kern 1pt} {\kern 1pt} {\kern 1pt} {\kern 1pt}{\left| {\bf{y}} \right|_2} = 1.
\end{align}
\end{subequations}
\emph{The optimal solution ${{\bf{y}}^*}$ is derived from the form of ${y^*} = c\sqrt {{\upsilon _n}} \chi _n^*$.}
\end{lemma}
 From the \textbf{Lemma 1}, the optimization problem in~\eqref{optimal3} is no longer an integer programming problem, nevertheless, this problem is still non-trivial as its feasible domain is non-convex. Then we use the SDP relaxation method to solve, that is, by removing certain constraints, the feasible set of the problem is converted into a positive semi-definite matrix cone.

Assume ${\bf{S}} = {\bf{y}}{{\bf{y}}^{\bf{T}}}$,~i.e., ${S_{n,n'}} = {y_n}{y_{n'}}$, then objective function in~\eqref{optimal3} can be converted to ${\bf{\mu }}_v^T{\bf{S}}{{\bf{\mu }}_{\bf{v}}}$ and the two constraints are respectively equivalent to $ {S_{nn'}}{\rm{ = 0,~ }}\forall n,n'$ if ${\Xi _{nn'}}{\rm{ = 1}}$ and $tr\left( S \right) = 1$. Because ${\bf{S}} = {\bf{y}}{{\bf{y}}^{\bf{T}}}$ and ${\bf{y}}$ is a matrix with dimension $N \times 1$, we can consider that $\left\{ {{\bf{S}} \in {\zeta ^{MN}}|{\bf{S}} \succ  = {\bf{O}},{\rm{rank}}({\bf{S}}) = 1} \right\}$, where $\zeta ^{MN}$ denotes the set of $MN \times MN$ real symmetric matrixes, and ${{\bf{S}} \succ  = {\bf{O}}}$ means ${\bf{S}}$ is positive semi-definite. Then we perform SDP relaxation on~\eqref {optimal3} by removing condition ${{\rm{rank}}({\bf{S}}) = 1}$, and get£¬
\begin{subequations}\label{SDP}
\begin{align}
&\theta \left( {\Xi ,v} \right) = \mathop {\max }\limits_{{\bf{S}} \succ  = {\bf{O}}} {\bf{\mu }}_v^T{\bf{S}}{{\bf{\mu }}_{\bf{v}}}\\
&{\rm{s}}{\rm{.t}}{\rm{.}}{\kern 1pt} {\kern 1pt} {\kern 1pt} {\kern 1pt} {\kern 1pt} {\kern 1pt} {\kern 1pt} {\kern 1pt} tr\left( S \right) = 1,\\
&{\kern 1pt} {\kern 1pt} {\kern 1pt} {\kern 1pt} {\kern 1pt} {\kern 1pt} {\kern 1pt} {\kern 1pt} {\kern 1pt} {\kern 1pt} {\kern 1pt} {\kern 1pt} {\kern 1pt} {\kern 1pt} {\kern 1pt} {\kern 1pt} {\kern 1pt} {\kern 1pt} {\kern 1pt}
{S_{nn'}}{\rm{ = 0, }}\forall n,n'{\kern 1pt} {\kern 1pt} {\kern 1pt} {\kern 1pt} {\rm{if}}{\kern 1pt} {\kern 1pt} {\kern 1pt} {\Xi _{nn'}}{\rm{ = 1}},
\end{align}
\end{subequations}
 which is a typical representation of the Lov¨¢sz number~\cite{Gr1988Geometric} in the graph theory.

By solving \eqref {SDP}, we can obtain the optimal ${{\bf{S}}^*}$. If $S_{nn}^* \ne 0$, then $\chi _n^* = 1$, and then check whether
the ``cache conflict" restriction relationship
of UTs in the same winner set meets the requirements. After multiple rounds of calculations, ${{\bf{S}}^*}$ may not be strictly equal to 0. At this time, a sufficiently small threshold can be set, such as ${10^{ - 5}}$, and $S_{nn}^*$ that is less than this threshold is very close to 0. Finally, we decompose ${{\bf{\chi }}^{\bf{*}}}$ into caching placement for different content chunks and get ${{\bf{x}}^{{{\left( 1 \right)}}}}{\bf{,~}}{{\bf{x}}^{{{\left( 2 \right)}}}}{\bf{,~}}...{\bf{,~}}{{\bf{x}}^{{{\left( M \right)}}}}$ and winner sets ${{\cal N}^{\left( 1 \right)}},~{{\cal N}^{\left( 2 \right)}},~...,~{{\cal N}^{\left( M \right)}}$.

\subsection{Pricing and Payment Strategy}
In the proposed multi-winner once auction model, the UTs in each set can be viewed as one bidder and the winner set is a set of UTs with the highest total bids. Hence, the pricing strategy of the proposed auction can be obtained using the same way as the pricing strategy of second-price sealed bid auction.
According to the second-price sealed bid auction mechanism, the pricing process of MOA is as follows:

1)~First, the problem~\eqref{optimal1} is solved to get the winner sets, ${\cal W}  =  \cup _{m = 1}^M{{\cal N}^{\left( m \right)}}$, that includes all content chunks.

2)~The winner sets, ${\cal W}$, are removed from the buyer set, and content chunk $m$ is auctioned again in the remaining loser set, ${\hat {\cal N}} = {\cal N}\backslash {\cal W}$, from which, we can calculate the maximum total actual revenue from~\eqref{optimal1}, denoted $U_{-{\cal W}}^{\left( m \right)*}$, for the winner set ${\cal N}^{\left( m \right)}$. In our pricing payment strategy, $U_{ - {\cal W}}^{\left( m \right)*}$ is set as the total price that the winners of content chunk $m$ need to pay to the BS.

3)~Then we propose a Nash bargaining solution (NBS) based payment strategy to determine the price $p_n^{\left( m \right)}$ for each UT in the winner set ${\cal N}^{\left( m \right)}$ to cache the content chunk $m$.


The goal of the proposed payment strategy in this paper is to balance the personal benefit of each UT in the winner set ${\cal N}^{\left( m \right)}$ as much as possible when the total price is equal to $U_{ - {\cal W}}^{\left( m \right)*}$.
We model the payment strategy as Nash bargaining game to allocate the price to be paid as evenly as possible in the winner set under the condition of guaranteeing their revenue. Nash bargaining game, as a cooperative game theory, is a popular strategy for fairly distributing resources among competing players~\cite{6784547}.
In the game, the players are UTs in the winner set ${\cal N}^{\left( m \right)}$ and player $n$, $n \in {\cal N}^{\left( m \right)}$, requires a minimal payoff $p_n^{\left( m \right)}$  to pay for content chunk $m$ with a constraint $\sum\limits_{n \in {\cal N}^{\left( m \right)}} {p_n^{\left( m \right)}}  = U_{ - {\cal W}}^{\left( m \right)*}$. Since ${v_n^{\left( m \right)}}$ has been known after the auction, the Nash bargaining problem can be formulated as

\begin{subequations}\label{price1}
\begin{align}
&\mathop {\max }\limits_{p_n^{\left( m \right)}} \prod\limits_{n \in {\cal N}^{\left( m \right)}} {\left( {v_n^{\left( m \right)} - p_n^{\left( m \right)}} \right)} \\
&{\rm{s}}{\rm{.t}}{\rm{.}}{\kern 1pt} {\kern 1pt} \sum\limits_{n \in {\cal N}^{\left( m \right)}} {p_n^{\left( m \right)} = U_{ - {\cal W}}^{\left( m \right)*}}, \\
&{\kern 1pt} {\kern 1pt} {\kern 1pt} {\kern 1pt} {\kern 1pt} {\kern 1pt} {\kern 1pt} {\kern 1pt} {\kern 1pt} {\kern 1pt} {\kern 1pt} {\kern 1pt} {\kern 1pt} {\kern 1pt} {\kern 1pt} {\kern 1pt} {\kern 1pt} {\kern 1pt} {\kern 1pt} {\kern 1pt} {\kern 1pt} {\kern 1pt} 0 \le p_n^{\left( m \right)} \le v_n^{\left( m \right)}.
\end{align}
\end{subequations}

Denote $q_n^{\left( m \right)} = v_n^{\left( m \right)} - p_n^{\left( m \right)}$, and $\sum\limits_{n \in {\cal N}^{\left( m \right)}} {v_n^{\left( m \right)} = {U^{{\left( m \right)}^*}}}$, then the problem~\eqref{price1} is equivalent to
\begin{subequations}\label{price2}
\begin{align}
&\mathop{\min }\limits_{q_n^{\left( m \right)}}  - \sum\limits_{n \in {\cal N}^{\left( m \right)}} {\ln q_n^{\left( m \right)}} \\
&{\rm{s}}{\rm{.t}}{\rm{.}}{\kern 1pt} {\kern 1pt} {\kern 1pt} {\kern 1pt} \sum\limits_{n \in {\cal N}^{\left( m \right)}} {q_n^{\left( m \right)} = {U^{{\left( m \right)}*}} - U_{ - {\cal W} }^{{\left( m \right)}*}}, \\
&{\kern 1pt} {\kern 1pt} {\kern 1pt} {\kern 1pt} {\kern 1pt} {\kern 1pt} {\kern 1pt} {\kern 1pt} {\kern 1pt} {\kern 1pt} {\kern 1pt} {\kern 1pt} {\kern 1pt} {\kern 1pt} {\kern 1pt} {\kern 1pt} {\kern 1pt} {\kern 1pt} {\kern 1pt} {\kern 1pt} {\kern 1pt}  0 \le p_n^{\left( m \right)} \le v_n^{\left( m \right)}.
\end{align}
\end{subequations}
According to~\cite{879352}, when $p_n^{\left( m \right)}$ is a concave upper-bounded function with convex support, there exists a unique and optimal NBS.
By solving~\eqref{price2} for the KKT conditions, we can get the price $p_n^{\left( m \right)}$ of UT $n$ in the winner set ${\cal N}^{\left( m \right)}$
that needs to pay to the BS.

Compared to the traditional VCG auction that may result in a low or even zero revenue in certain circumstances, the revenue of UTs in the proposed auction model is more assured so the UTs are motivated to participate in the auction. At the same time, if the losers want to defeat the winners by lying, they will need to bid more than $U_{ - {\cal W} }^{{\left( m \right)}*}$. Since the price for winning the auction of the contents is greater than the actual value of the content for the losers, the losers will have no incentive to lie, thereby eliminating the possibility of collusion between the UTs. However, as long as there is a manageable gap between the price of a winner and the bid of losers, there is still the possibility that the winner will sublease the content chunk to the losers. In the event of a sublease, the obtained caching placement will no longer be the best strategy and will cause losses for caching performance. Therefore, we need to further to avoid the occurrence of sub-leasing in~\eqref{price2}.

We assume that some of the UTs in the winner set ${\tilde {\cal N}^{\left( m \right)}}$ sublease their content chunks to the losers ${\tilde {\cal L}^{\left( m \right)}}$, where ${\tilde {\cal N}^{\left( m \right)}} \subseteq {{\cal N}^{\left( m \right)}}$, ${\tilde {\cal L}^{\left( m \right)}} \subseteq {\hat {\cal N}}$. From the above auction and pricing process, we can see that sub-leasing can only occur when the sub-lease price is higher than the price paid by some of the winners and lower than the price of some of the losers. The necessary prerequisite for sub-leasing  is $\sum\limits_{n \in {{\tilde {\cal N}}^{\left( m \right)}}} {p_n^{\left( m \right)} < \sum\limits_{n \in {\tilde {\cal L}^{\left( m \right)}}} {v_n^{\left( m \right)}} }$. As long as this premise is avoided, sub-leasing will not happen. It should be noted that potential sub-leasing UTs can only successfully lease content chunks when there is no cache conflict among UTs in the winner set. We define ${{\cal L}^{\left( m \right)}} \buildrel \Delta \over = \{ n \in {\hat {\cal N}}|{E_{nn'}} = 0,~\forall n' \in {{\cal N}^{\left( m \right)}}\backslash {\tilde {\cal N}^{\left( m \right)}}\} $ to denote these UTs, then there should be $\sum\limits_{n \in {{\tilde {\cal N}}^{\left( m \right)}}} {p_n^{\left( m \right)} \ge \mathop {\max }\limits_{{{\tilde {\cal L}}^{\left( m \right)}} \subseteq {{\cal L}^{\left( m \right)}}} \sum\limits_{n \in {{\cal L}^{\left( m \right)}}'} {v_n^{\left( m \right)}} } $, where $\mathop {\max }\limits_{{{\tilde {\cal L}}^{\left( m \right)}} \subseteq {{\cal L}^{\left( m \right)}}} \sum\limits_{n \in {{\tilde {\cal L}}^{\left( m \right)}}} {v_n^{\left( m \right)}} $ is the maximum social welfare $U_{{{\cal L}^{\left( m \right)}}}^*$ in UT set ${\tilde {\cal L}^{\left( m \right)}}$, which is obtained by re-auctioning content chunk $m$. Therefore, we should restrict~\eqref{price1} as
\begin{subequations}\label{price3}
\begin{align}
&\mathop {\max }\limits_{p_n^{\left( m \right)}} \prod\limits_{n \in {{\cal N}^{\left( m \right)}}} {\left( {v_n^{\left( m \right)} - p_n^{\left( m \right)}} \right)} \\
&{\rm{s}}{\rm{.t}}{\rm{.}}{\kern 1pt} {\kern 1pt} {\kern 1pt} {\kern 1pt} \sum\limits_{n \in {{\tilde {\cal N}}^{\left( m \right)}}} {p_n^{\left( m \right)} \ge U_{{{\cal L}^{\left( m \right)}}}^{{\left( m \right)}*},\forall {{\tilde {\cal N}}^{\left( m \right)}} \in {{\cal N}^{\left( m \right)}}} \\
&{\kern 1pt} {\kern 1pt} {\kern 1pt} {\kern 1pt} {\kern 1pt} {\kern 1pt} {\kern 1pt} {\kern 1pt} {\kern 1pt} {\kern 1pt} {\kern 1pt} {\kern 1pt} {\kern 1pt} {\kern 1pt} {\kern 1pt} {\kern 1pt} {\kern 1pt} {\kern 1pt} {\kern 1pt} {\kern 1pt} {\kern 1pt}{\kern 1pt} {\kern 1pt} 0 \le p_n^{\left( m \right)} \le v_n^{\left( m \right)}.
\end{align}
\end{subequations}
When ${\tilde {\cal N}^{\left( m \right)}} = {{\cal N}^{\left( m \right)}}$, the constraint is $\sum\limits_{n \in {{\cal N}^{\left( m \right)}}} {p_n^{\left( m \right)}}  \ge U_{ - {\cal W} }^*$, which means the constraint in~\eqref{price1} is a special case in~\eqref{price3}. In addition to the case of ${\tilde {\cal N}^{\left( m \right)}} = \emptyset $, where ${{\cal N}^{\left( m \right)}}$  has ${2^{\left| {{{\cal N}^{\left( m \right)}}} \right|}} - 1$  subsets. To finally determine the constraints, it is necessary to calculate the maximum social welfare for the optimal caching placement of these subsets.

The process of MOA based caching (MOAC) placement algorithm is summarized in \textbf{Algorithm 1}.
\vspace{-0.5em}
\begin{algorithm}
\small
\caption{MOAC algorithm}
\label{MOAC}
\begin{algorithmic}
\STATE 1: - \textbf{\textit{Step 1: Initialization}}
\STATE 2: Calculate the content sharing profit of content chunk $m$ for UT $n$ $v_n^{\left( m \right)}\left( {m \in [1,M],n \in [1,N]} \right)$ through~\eqref{value} to obtain ${\bf{\upsilon }} = [{{\bf{v}}^{\left( 1 \right)}},~{{\bf{v}}^{\left( 2 \right)}},~...,~{{\bf{v}}^{\left( M \right)}}]$;
\STATE 3: Initial ${\bf{\mu }} = {[\sqrt {{\upsilon _1}} ,\sqrt {{\upsilon _2}} ,...,\sqrt {{\upsilon _{MN}}} ]^T}_{}$, ${\bf{y}} = \left[ {c\sqrt {{\upsilon _1}} {\chi _1},c\sqrt {{\upsilon _2}} {\chi _2},...,c\sqrt {{\upsilon _{MN}}} {\chi _{MN}}} \right]$, ${\bf{S}} = {\bf{y}}{{\bf{y}}^{\bf{T}}}$, and $\Xi $ in~\eqref{optimal2};
\STATE 4: - \textbf{\textit{Step 2: Determine the caching placement}}
\STATE 5: Solve the problem~\eqref{SDP}, get ${{\bf{S}}^*}$;
\STATE 6: If ${\bf{S}}_{nn}^{\rm{*}} \ne 0$, $\chi _n^* = 1$, get ${{\bf{\chi }}^{\bf{*}}}$;
\STATE 7: Decompose ${{\bf{\chi }}^{\bf{*}}}$  into caching placement for different content chunks, get ${\bf{\chi }}{\rm{ = }}\left[ {{{\bf{x}}^{\left( 1 \right)}},{{\bf{x}}^{\left( 2 \right)}},...,{{\bf{x}}^{\left( M \right)}}} \right]$ and winner sets ${\cal W} =  \cup _{m = 1}^M{{\cal N}^{\left( m \right)}}$;
\STATE 8: - \textbf{\textit{Step 3: Pricing and payment strategy}}
\STATE 9: Solve problem~\eqref{price2} for content chunk $m$ $\left( {m \in [1,M]} \right)$ to obtain price ${\bf{P = }}{\left[ {{{\bf{p}}^{{{\left( 1 \right)}}}}{\bf{,~}}{{\bf{p}}^{{{\left( 2 \right)}}}}{\bf{,~}}...{\bf{,~}}{{\bf{p}}^{{{\left( M \right)}}}}} \right]^T}$;
\STATE 10: - \textbf{\textit{Step 4: End of the algorithm}}
\end{algorithmic}
\end{algorithm}
\vspace{-0.5em}
\subsection{Discussions}
In this part, we discuss the optimality, convergence, complexity and scalability of the proposed MOAC. Besides, we summarize that the MOAC satisfy the three auction principles mentioned in the proposed auction model.

\subsubsection{Optimality}
Since we relax the original optimization problem by removing the condition ${{\rm{rank}}({\bf{S}}) = 1}$ to obtain the approximate optimal solution of caching placement, problem~\eqref{SDP} provides an upper bound for original optimization problem~\eqref{optimal1}. As one of the constraints, the rank of matrix ${\bf{E}}$ will have an influence on the rank of solved matrix ${\bf{S}}$, which in turn affects the optimality of the SDP relaxation. The rank of matrix ${\bf{E}}$ is related to the value of $\gamma$. When $\gamma$ is close to the limit value of 1 or 0, the rank of ${\bf{E}}$ is small, so that the influence of ${\bf{E}}$ on the optimality of SDP is very small. In this case, we can almost obtain the optimal solution. Conversely, when other value of $\gamma$ is given, the SDP relaxation obtains the upper bound of original optimization with a larger rank of matrix ${\bf{E}}$. In such case, there is a tiny  performance gap between  the optimal  solution and the relaxed solution.
\begin{remark}\label{remark2}
If the optimal solution ${{\bf{S}}^*}$ can be decomposed into ${{\bf{S}}^{\bf{*}}} = {{\bf{y}}^{\bf{*}}}{\left( {{{\bf{y}}^{\bf{*}}}} \right)^T}$, it means that ${{\bf{S}}^*}$ is in the feasible set of the original optimization problem; otherwise, $\theta \left( {\Xi ,v} \right)$ is the upper bound that the original optimization problem cannot reach. Nevertheless, based on our simulation results in Section IV, the relaxed approximate optimization problem has more than 99\% of possible solutions in the feasible set.
\end{remark}
\subsubsection{Convergence}
We use the primal-dual interior point method to solve the SDP problem in~\eqref{SDP}. Studies have shown that the primal-dual interior point method~\cite{Yamashita2003SDPARA} is one of the most effective algorithms for solving SDP problem. This algorithm has good convergence and can guarantee accurate and optimal solutions. We ensure that the relative gap between the primal and the dual solution is less than $1.49\times {10^{ - 8}}$ in our simulation. The convergence speed of the algorithm is related to the complexity determined by the degree of freedom of the matrix $\bf{S}$.

\subsubsection{Complexity}
Since we regard the optimization problem as an once-auction with $MN$ bidders (UTs) participating, we need to optimize the problem~\eqref{SDP} over ${{\bf{S}} \in {\zeta ^{MN}}}$, which has $\frac{1}{2}MN\left( {MN + 1} \right)$ degrees of freedom. The computational complexity of MOAC is $O({M^2}{N^2})$, which mean that the calculation time of the MOAC will increase significantly as the number of content chunks increases.
\subsubsection{Scalability}
In our caching model, we assume that each UT can cache one content chunk with size of $s$. In fact, the proposed auction model and MOAC algorithm can be generalized as UT $n$ caching $h_n$ content chunks with cache store size of ${h_n}s$. In this case, UT $n$ will be regarded as $h_n$ UTs, and the bidder set in the auction can be extended as
 ${{\bar {\cal N}}}= \{ {\bar {n}}|  {\bar {n}} \in {{\cal N}_n}, {n \in {\cal N}}  \}$, where ${{\cal N}_n}= \{n_{1}, n_{2},..., n_{{h_n}}|{n \in {\cal N}}\}$.
Correspondingly, the caching placement can be obtained via problem~\eqref{optimal1} in the new bidder set ${{\bar {\cal N}}}$.
\subsubsection{Auction Properties}
Since we consider the revenue of all the UTs  as social welfare, so the social efficiency can be ensured in~\eqref{optimal1}. At the same time, only when the UT's revenue for caching a chunk is greater than 0, it will participate in the auction. This assumption is in line with individual rationality of each UT. Finally, the second-price sealed bid auction we discuss in the pricing payment strategy can guarantee the MOAC's truthfulness.
\vspace{-0.5em}
\section{Multi-Winner Repeated Auction based Caching Placement}
In the practical cellular networks, the content number is extremely large, the MOAC takes a very long time to get the caching placement results, which has limited significance in practical scenarios. In this section, we propose a simplified caching placement algorithm with low complexity.

Since the popularity of the contents in the mobile networks follows the Zipf distribution~\cite{Breslau2002Web}, only a small portion of the content chunks are quested by most of the UTs in a cell. Obviously, it is only necessary to auction high-popularity contents and ignore low-popularity ones. Therefore, we propose a heuristic MRAC based on the popularity of content chunks. In the proposed MRAC, the content chunks are sorted according to the local popularity of the contents in one cell, then the content chunks are auctioned one by one from the highest popularity until no UT in the cell is willing to continue participating in the auctions.

First, the local popularity ${F_m} = \frac{{\sum\limits_{n = 1}^N {f_n^{\left( m \right)}} }}{{\sum\limits_{m = 1}^M {\sum\limits_{n = 1}^N {f_n^{\left( m \right)}} } }}$ of chunk $m$ is calculated according to $f_n^{\left( m \right)}$. Then the content chunks are ranked according to ${F_m}$ from highest to lowest. The most popular content chunk is denoted as 1, and the content chunk with popularity $i$ is denoted as $i$, $1 \le i \le M$. The auction starts from the most popular content chunk one by one. Since each UT can only cache one content chunk, UT $n$ will not participate in the next round auction if its cache space has been occupied. In the $i^{\rm{th}}$ round auction, the participant UTs is defined as
${n \in {{\cal L}^{\left( i \right)}}}$, then the caching placement problem as follows,
\begin{subequations}\label{optimal4}
\begin{align}
&\mathop {\max }\limits_{{{\bf{x}}^{\left( i \right)}}} \sum\limits_{n = 1}^N {v_n^{\left( i \right)}x_n^{\left( i \right)}} \\
\label{18b}
&{\rm{s}}{\rm{.t}}{\rm{.}}{\kern 1pt} {\kern 1pt} {\kern 1pt} {\kern 1pt} x_n^{\left( i \right)} + x_{n'}^{\left( i \right)} \le 1,{\kern 1pt} {\kern 1pt} {\kern 1pt} {\kern 1pt} \forall n,n'{\kern 1pt} {\kern 1pt} {\kern 1pt} {\kern 1pt} \rm{if}{\kern 1pt} {\kern 1pt} {\kern 1pt} {\kern 1pt} {E_{n,n'}} = 1,\\
\label{18c}
&{\kern 1pt}{\kern 1pt}{\kern 1pt}{\kern 1pt}{\kern 1pt}{\kern 1pt}{\kern 1pt}{\kern 1pt} {\kern 1pt} {\kern 1pt} {\kern 1pt} {\kern 1pt} {\kern 1pt} {\kern 1pt}{\kern 1pt}{\kern 1pt}{\kern 1pt}{\kern 1pt}{\kern 1pt}{\kern 1pt}x_n^{\left( i \right)} = \left\{{\rm{0}}, {\rm{1}} \right\}, {\kern 1pt}{\kern 1pt}{\kern 1pt}{\kern 1pt}
{n \in {{\cal L}^{\left( i \right)}}},
\end{align}
\end{subequations}
where~\eqref{18b} denotes the ``cache conflict" restrict relationship of UT $n$ and UT $n'$, and~\eqref{18c} constrains that $x_n^{\left( i \right)}$ is a binary variable.
In fact, problem~\eqref{optimal4} can be regarded as a special case of problem~\eqref{optimal1} with $M=1$. The process of MRAC is given in \textbf{Algorithm 2}.
\begin{algorithm}
\small
\caption{MRAC algorithm}
\label{MRAC}
\begin{algorithmic}
\STATE 1: - \textbf{\textit{Step 1: Initialization}}
\STATE 2: Rank the content chunks according to the local popularity ${F_m}$ from the highest to the lowest, and the content chunk with popularity $i$ is denoted as $i$, $1 \le i \le M$;
\STATE 3: - \textbf{\textit{Step 2: content chunk $i$ auction process}}
\STATE 4: In the $i^{\rm{th}}$ round of auction, obtain the winner set ${{\cal N}^{\left( i \right)}}$ and caching placement for the content chunk $i$ from the solution of problem~\eqref{optimal4};
\STATE 5:  Determine the price of content chunk $i$ using the pricing payment strategy described in~\eqref{price1}-\eqref{price2}, the price ${{\bf{p}}^{\left( i \right)}} = {[p_1^{\left( i \right)},p_2^{\left( i \right)},...,p_N^{\left( i \right)}]^T}$ based on Nash bargaining game is obtained.
\STATE 6: Move the winner set ${{\cal N}^{\left( i \right)}}$ out from the UTs set, and the participant UTs set of $(i+1)^{\rm{th}}$ round is ${{\cal L}^{\left( i+1 \right)}} = {\cal N} \backslash \sum\limits_{j = 1}^i {{{\cal N}^{\left( j \right)}}} $.
\STATE 7: - \textbf{\textit{Step 3: Judgement }}
\STATE 8: \textbf{if} ${{\cal L}^{\left( i+1 \right)}}$ is not an empty set \textbf{then}
\STATE 9: $i=i+1$;
\STATE 10: Goto \textbf{\textit{Step 2}};
\STATE 11: \textbf{else}
\STATE 12: End of the algorithm;
\STATE 13: \textbf{end if}
\end{algorithmic}

\end{algorithm}

\begin{figure} [t!]
\centering
\includegraphics[width=3.5in]{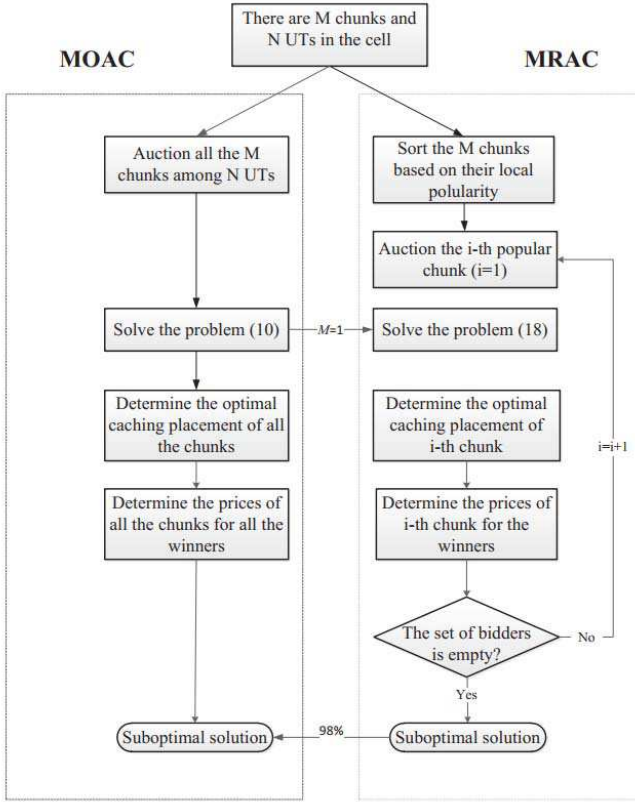}
 \caption{Comparison of MOAC and MRAC algorithms.}
 \label{algorithms}
 \vspace{-1em}
\end{figure}

We compare the proposed \textbf{Algorithm 1} (MOAC) and \textbf{Algorithm 2} (MRAC) in~Fig.~\ref{algorithms}. MOAC auctions all the contents for all the UTs by one time auction, which gives the near optimal caching placement solution. In contrast, MRAC auctions the contents one by one and the final caching placement solution is obtained after multiple auctions. MRAC can be seen as a simplified algorithm of MOAC since the problem~\eqref{optimal4} of MRAC is a special case of problem~\eqref{optimal1} when $M=1$. The complexity of MRAC is much lower than that of MOAC. In MRAC, we only need to solve the problem over ${\bf{S}}$ with degrees of $\frac{1}{2}N(N + 1)$ in every round of auction. Due to the limited UTs cache spaces, the number of auction rounds based on content chunks popularity rankings is limited and does not increase as the number of content increases. So we can consider the computational complexity reduced from $O({M^2}{N^2})$ to $O({N^2})$.

\begin{remark}\label{remark3}
MOAC and MRAC get similar performance under the condition that the popularity of content chunks are Zipf distribution but the algorithm complexity of MRAC is very low. In MRAC, the most important UTs will cache the most popular content chunks, which is consistent with the initial goal of caching placement in MOAC. The performance gap between MRAC and MOAC is very small. Simulation results also show that, MRAC gets 98\% caching performance of MOAC while the calculation time of MRAC is greatly reduced with larger number of UTs. So we can apply MRAC in the practical edge caching in D2D cellular networks.
\end{remark}

In order to implement the proposed MOAC and MRAC algorithms, it is a feasible solution to make a centralized agent entity with powerful computation ability, such as a functional entity deployed in the macro BS, to host the auction for content caching UTs and other cache management function, such as the cache update function and cache consistency maintenance function\footnote{In a dynamic network with user movements, a learning based distributed caching placement algorithm may be more feasible in practical scenarios.}. In this way, the mobile users can report the control information, such as, the encounter probability between UTs, and the average transmission rate between UTs, to this centralized agent entity.
Based on the proposed auction model, we formulate the optimization problem and seek for an upper bound for the content caching placement. The computational complexity of the suboptimal solution is high for individual UT.

Since we assume that there is a centralized agent entity for cache management, it is nature to implement the cache consistency maintenance in this centralized agent entity. Some state-of-art cache consistency maintenance algorithms have been proposed for wireless mobile environments, such as~\cite{CacheCon}. The centralized agent entity records all cached contents for each UT. When the cached contents are expiration during UTs' movement, it sends an invalidation report (IR) to that particular UT. The cache consistency maintenance is out of the scope of this paper.
Meanwhile, the proposed algorithms are periodically executed in a predefined time interval, such as, one hour or one day. Consequently, the cache of UT can be updated under the control of the centralized agent entity.

In this paper, besides the general control overhead for D2D communications in cellular networks, the control overhead of the proposed algorithms is mainly derived from the submission of the UTs¡¯ bids during the auction and delivery the auction results to the UTs according to the proposed algorithms. Therefore, our control overhead is proportional to the number of UTs participating in the auction. Since the information exchange between the centralized agent entity and the UTs occurs in the off-peak period, the performance impact on the peak period can be negligible compared to the calculation cost of the centralized agent entity.

\vspace{-0.8em}
\section{Numerical Results}
In this section, the system performance of the proposed caching placement algorithms are verified by Matlab simulation.
\vspace{-0.5em}
\subsection{Simulation Settings}
In the simulation, a macro BS is deployed at the center of the cell and $N$ UTs are randomly distributed in the cell and can communicate with any neighbor UTs in within the coverage.
The clustered random  model is used to model the UTs' movement in the macrocell with a coverage radius $R$. We assume that there are $C$ clusters within this coverage, and the radius of the middle point of each cluster is $30$~m. We assume that there are $H$ home-points distributed within the radius of each cluster, and each home-point is shared by ${n_h}$ UTs. Therefore, the number of UTs in our simulated cellular network is $N = CH{n_h}$.
In the simulation, the parameter setting used for D2D communication is from the Technical Report of 3GPP~\cite{3gpp}, the system bandwidth of D2D communication is 10 MHz uplink and 10 MHz downlink for FDD, and the indoor to indoor channel model is as defined in~\cite{3gpp}, including the pathloss, shadowing, and the fast fading.
The detailed simulation parameters are given in Table II.

\begin{table}[!htbp]
\small
  \begin{center}
      \caption{Simulation Parameters}
    \begin{tabular}{|p{4cm}<{\centering}|l|}
      \hline
      \textbf{Parameter} & \textbf{Value}  \\       \hline
      Carrier frequency & 2 GHz \\ \hline
      Radio bandwidth & 20 MHz\\ \hline
      Backhaul data rate & 1.5 Mbps \\ \hline
      D2D transmit power $P_{n'}^{tx}$ & 23 dBm  \\ \hline
      BS transmit power $P_{BS}^{tx}$ & 43 dBm \\ \hline
      Pathloss from BS to UT & 37.6log10(d[km])+128.1 dB    \\  \hline
      Pathloss of D2D channel & 40log10(d[km])+148 dB   \\ \hline
      Noise power spectral density & -174 dBm/Hz   \\ \hline
      $K$ & -50 dBm  \\ \hline
      $\vartheta $ & $10^{-6}$ \\ \hline
      $\zeta$ & 1 \\ \hline
     \end{tabular}
  \end{center}
  \vspace{-0.5em}
\end{table}

The popularity of $M$ contents follows a \emph{Zipf}-like distribution~\cite{Breslau2002Web} with parameter $\alpha$. When the contents are indexed in descending order of popularity, the popularity probability of the $m^{\rm{th}}$ content is
\begin{align}
{p_m} = {m^{ - \alpha }}/\sum\limits_{j = 1}^M {{j^{ - \alpha }}},
\end{align}
the popularity distribution is more skewed with larger $\alpha$.
The size of each content chunk $s$ is set to one M~bytes. We use ${f_n^{\left( m \right)}}$ to model the data request of UT $n$ for the content $m$, which is assumed to be a constant during the UT movement. For UT $n$, we assume that $\sum\limits_{m = 1}^M {{f_n^{\left( m \right)}}}  = 1$.
The BS makes the caching placement decision with the statistical knowledge of content preference and movement position of UTs.
\vspace{-0.5em}
\subsection{Simulation Results}
\begin{figure} [t!]
\centering
\includegraphics[width= 3.65in, height=2.56in]{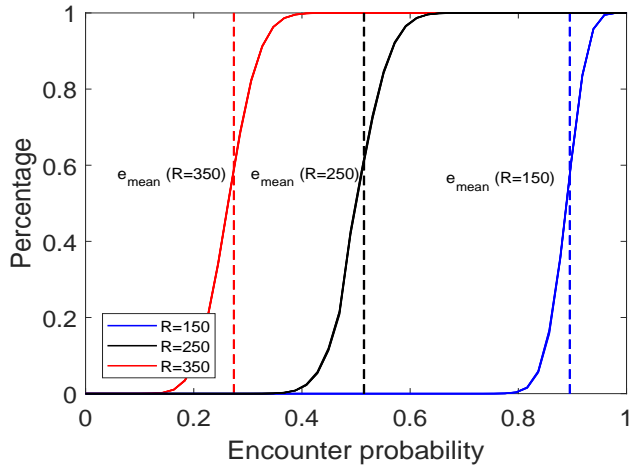}
 \caption{Cumulative distribution of the encounter probabilities.}\label{cdf}
  \vspace{-0.5em}
\end{figure}

\begin{figure} [t!]
\centering
\includegraphics[width= 3.7in, height=2.6in]{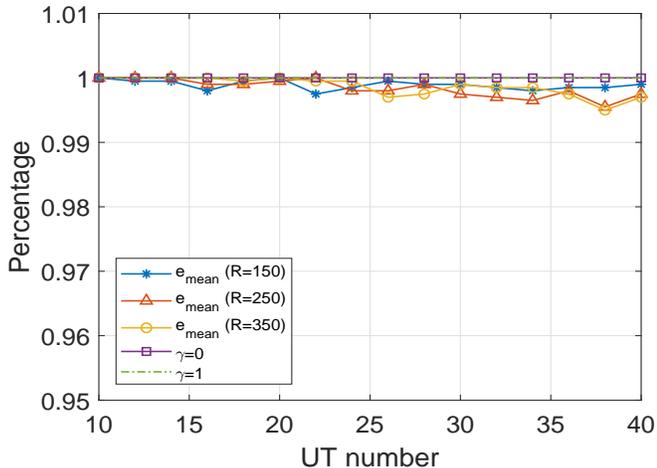}
 \caption{Proportion of the optimal caching placement solutions obtained by MOAC.}\label{opt}
  \vspace{-1em}
\end{figure}

We first demonstrate the effectiveness of the proposed MOAC with SDP relaxation. We set $C=2$ and vary  $N$ from 10 to 40. It can be seen from our previous discussion that the value of encounter probability threshold $\gamma$ is related to the encounter probabilities between UTs, which determines the ``cache conflict"  matrix and also affects the optimality of our SDP relaxation. Therefore, $\gamma$ should choose an appropriate value according to the distribution of the encounter probabilities between UTs. In Fig.~\ref{cdf}, we demonstrate the encounter probabilities between UTs in the same cluster with BS coverage radius of 150~m, 250~m and 350~m. Obviously, the smaller the coverage radius of the BS is, the higher mean value of the encounter probabilities is. The cumulative distribution results also show that, within the same size coverage, the encounter probability between two specific UTs in the same cluster will be concentrated in a certain span of about 0.2. Based on this statistical result, we choose the  mean value of in-cluster encounter probabilities ${e_{mean}}$ as the value of $\gamma$,
\begin{align}
\small
{e_{mean}} = \frac{1}{C}\sum\limits_{c = 1}^C {\frac{1}{{\left| {{{\bf{e}}_c}} \right|}}\sum\limits_{{e_{n,n'}} \in {{\bf{e}}_c}} {{e_{n,n'}}} },
\end{align}
where ${\bf{e}}_c$ denotes the encounter probabilities matrix between UTs within cluster $c$ and ${\left| {{{\bf{e}}_c}} \right|}$ denotes the number of encounter probabilities in ${\bf{e}}_c$. In this case, the value of $\gamma$ is variable depending on the BS coverage radius and UTs movement model. In Fig.~\ref{opt}, when we take ${\gamma  = {e_{mean}}}$, the caching placement obtained by MOAC is compared with the optimal caching placement obtained from the exhaustive search method with 1000 rounds of calculations for varying UT numbers. The ratio of solutions obtained by MOAC with SDP relaxation to the optimal solutions by exhaustive search is always above 99\%, which demonstrates the~\textbf{{Remark~\ref{remark2}}} we have provided in section III. Therefore, it can be seen that the proposed MOAC based on SDP relaxation is feasible and effective.

\begin{figure} [t]
\centering
\includegraphics[width= 3.65in, height=2.56in]{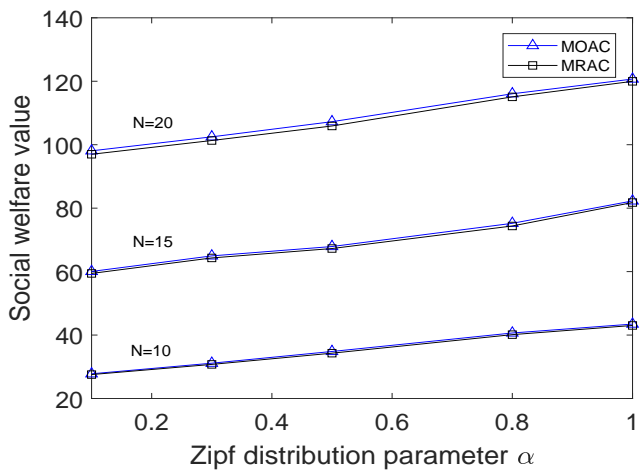}
 \caption{Social welfare value comparison of MOAC and MRAC.}
 \label{MOA-MRA}
  \vspace{-0.5em}
\end{figure}

\begin{figure} [t]
\centering
\includegraphics[width= 3.65in, height=2.56in]{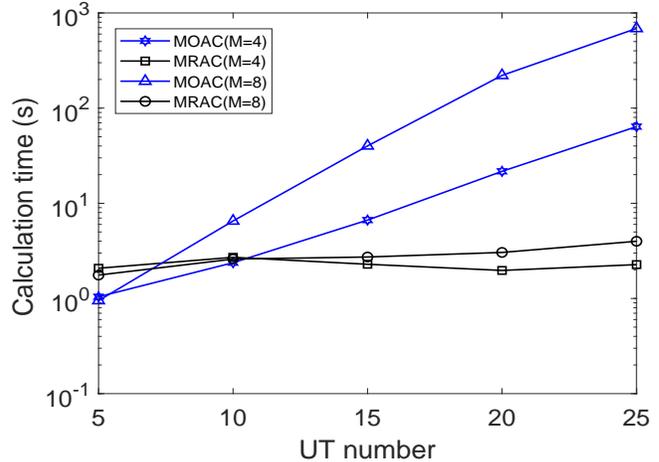}
 \caption{Calculation time comparison of MOAC and MRAC.}
 \label{time}
  \vspace{-0.5em}
\end{figure}

\begin{figure} [tb]
\centering
\includegraphics[width= 3.65in, height=2.56in]{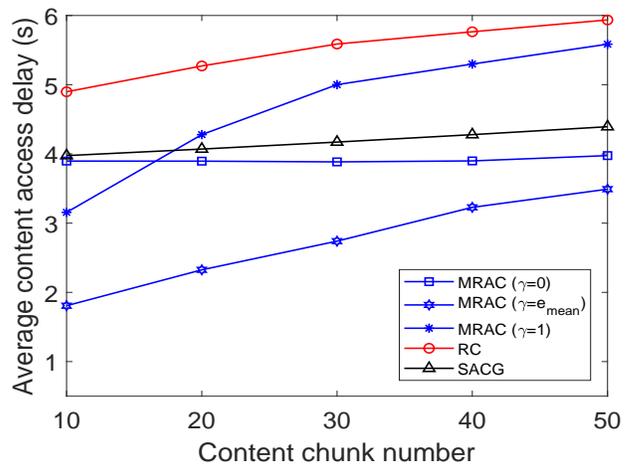}
 \caption{Average content access delay comparison with varying number of content chunks.}
 \label{delay}
 \vspace{-1.5em}
\end{figure}

\begin{figure} [tb]
\centering
\includegraphics[width= 3.65in, height=2.56in]{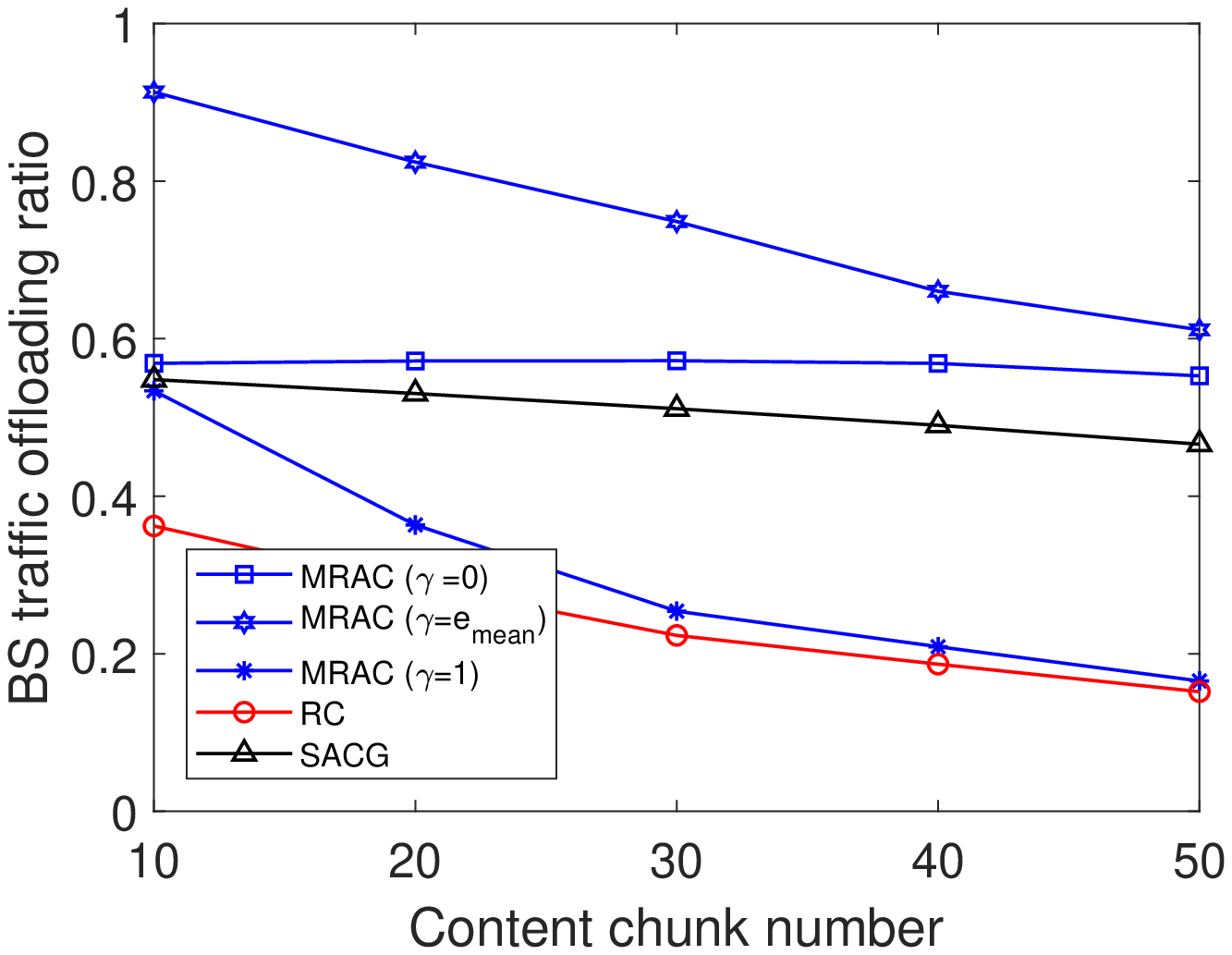}
 \caption{BS traffic offloading ratio comparison with varying number of content chunks.}
 \label{offloading}
   \vspace{-0.5cm}
\end{figure}

Then we compare the performance of MOAC and MRAC in terms of social welfare value and calculation time complexity.
Because MRAC sorts and auctions content chunks according to the popularity, we consider the influence of the zipf distribution parameter of the content chunks popularity on the performance of MOAC and MRAC in the simulation.
Fig.~\ref{MOA-MRA} compares the social welfare values of MOAC and MRAC when the number of UTs is 10, 15 and 20 respectively. The social welfare value of MRAC is almost the same as that of MOAC in the three cases, with only 2\% difference.
As the zipf distribution parameter increases from 0.1 to 1, the average social welfare value ratio of MRAC to MOAC decreases from 99.41\% to 99.16\%, which means that the zipf distribution parameter has little effect on the performance gap between MRAC and  MOAC.
Fig.~\ref{time} compares the calculation time of MOAC and MRAC with 4 content chunks and 8 content chunks. The simulation results show that, i)~when the number of content chunks increases form 4 to 8, the calculation time of MOAC increases sharply,
in contrast, the calculation time of MRAC increases mildly; ii)~the calculation time of MOAC increases quickly  with the increasing of UT numbers, but that of MRAC has little changes.
Fig.~\ref{MOA-MRA} and Fig.~\ref{time} demonstrate the~\textbf{{Remark~\ref{remark3} }}we have provided in section IV. Therefore, the proposed MRAC is more feasible in the practical scenario with a large number of contents.

Next, we compare the caching performance of the proposed MRAC with the social-aware caching game (SAGG)~\cite{Zhu2017Social} and  random caching (RC).
The criteria of the caching performance considered in this paper are average content access delay and the BS traffic offloading ratio.
In the simulation results, the average content access delay of $N$ UTs for $M$ chunks is
\begin{align}\label{delay1}
\scriptsize
 D = {\mathbb{E}_n}\left\{ {\sum\limits_{m = 1}^M {f_n^{\left( m \right)}\left[ {\frac{s}{{r_{n,n'}^{\min }}}\theta  + \left( {\frac{s}{{{r_{n,BS}}}} + \frac{s}{{{r_{back}}}}} \right)\left( {1 - \theta } \right)} \right]} } \right\}
\end{align}
where ${r_{n,n'}^{\min }}$ denotes the minimum data rate of D2D communication between UT $n$ and $n'$
($n' \in {\cal N}_n^{\left( m \right)}$ and ${\cal N}_n^{\left( m \right)} = \{ n'|x_{n'}^{\left( m \right)} = 1,n' \in {{\cal N}_n}\} $
during the content delivery, ${{r_{back}}}$ denotes the data rate of the backhaul link of the BS and $\theta  = 1$ if ${N_d} \ne \emptyset$, which means that chunk $m$ is shared by D2D communication in the cell, else $\theta  = 0$, which is the case of BS transmission. $\mathbb{E}_n$ indicates the average access time for all UTs to obtain all their required chunks.
The BS traffic offloading ratio is calculated as
\begin{align}\label{Offloading}
\small
O = \frac{{\sum\limits_{m = 1}^M {\sum\limits_{n = 1}^N {f_n^{\left( m \right)}x_n^{\left( m \right)}} } }}{{\sum\limits_{m = 1}^M {\sum\limits_{n = 1}^N {f_n^{\left( m \right)}} } }}.
\end{align}
The larger value of $O$ means more traffic are delivered by D2D communications.

In the simulation, we assume that coverage radius $R=250~m$, the number of UTs $N=40$, and the zipf distribution parameter $\alpha = 1$. We simulate the performance of MRAC when $\gamma$ is equal to 0, 1 and ${e_{mean}}$, respectively. As we discussed before, when $\gamma=1$, there is no ``cache conflict" between the UTs, which will cause almost all UTs to cache the same most popular chunk. When $\gamma=0$, the ``cache conflict" condition between UTs is very strict, that is, only one chunk can be cached in the entire macrocell region regardless of whether there is D2D communication capability between UTs.

Fig.~\ref{delay} compares the average content access delay for different numbers of content chunks. As shown in Fig.~\ref{delay}, no matter which algorithm is used for proactive caching at off-peak hours, the average content access delay for content delivery during peak hours can be reduced since the average content access delay by the BS transmission without caching during the peak period is 6.3 s in our simulation.
When $\gamma=1$, as the number of chunks increases, the average content access delay of MRAC increases rapidly, which is gradually higher than SACG and close to RC. When $\gamma=0$, the average content access delay of MRAC is very close to SACG. When $\gamma={e_{mean}}$, the average content access delay of MRAC can be significantly reduced. When the number of contents reaches 50, the average content access delay of MRAC is 79.46\% and 58.81\% of RC and SACG, respectively.
From Fig.~\ref{offloading}, when $\gamma={e_{mean}}$, MRAC can effectively reduce the traffic load of the BS during peak hours.
When the number of contents reaches 50, MRAC's BS traffic offloading is about 4.02 times and 1.31 times of RC and SACG, respectively.
Since the cache space is limited, the BS traffic offloading decreases as the number of content chunks increases. Similar as the performance of average content access delay, when $\gamma=1$ and 0, the BS traffic offloading performance of MRAC is close to RC and SACG, respectively.
The results of Fig.~\ref{delay} and Fig.~\ref{offloading} verify that the proposed MRAC can effectively reduce the average content access delay of the UTs and the traffic load of the BS in peak hours, which demonstrates the {\bf{ {Remark~\ref{remark1}}}} we provided in section III.

\begin{figure} [t]
 \vspace{-0.5em}
\centering
\includegraphics[width= 3.63in, height=2.5in]{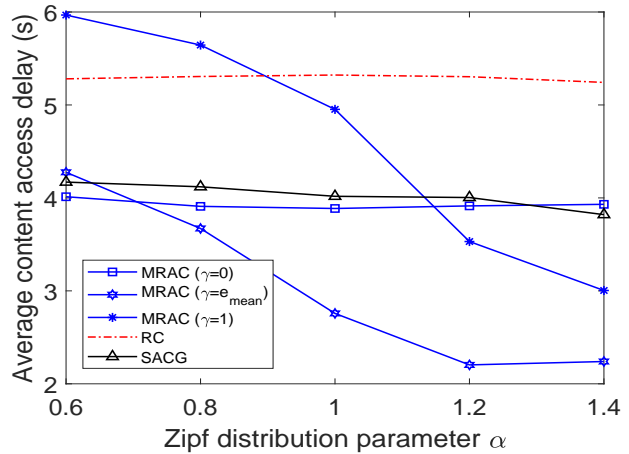}
\caption{Average content access delay comparison with varying zipf distribution parameter.}
 \vspace{-1em}
 \label{delay_zipf}

\end{figure}

\begin{figure} [t]
 \vspace{-0.5em}
\centering
\includegraphics[width= 3.63in, height=2.5in]{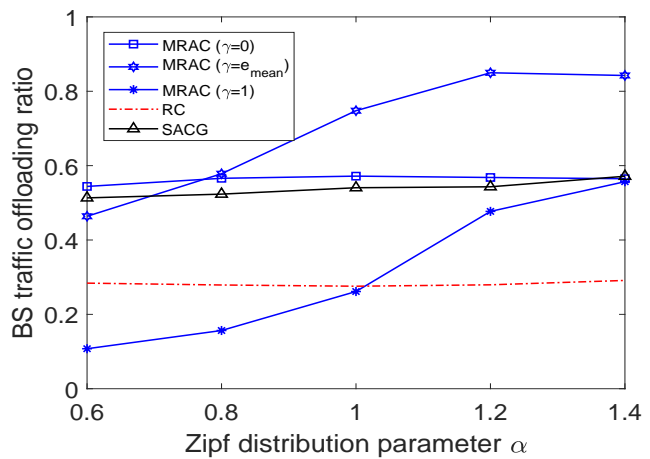}
\caption{BS traffic offloading ratio with varying zipf distribution parameter.}
 \vspace{-1em}
 \label{offloading_zipf}

\end{figure}

Finally, we consider the impact of the zipf distribution parameter $\alpha$ on the performance of the caching placement algorithms. We set $M = 30$, $N =40$, and vary $\alpha$ from 0.6 to 1.4. As shown in Fig.~\ref{delay_zipf} and Fig.~\ref{offloading_zipf}, as $\alpha$ increases, the average content access delay of MRAC decreases from 4.27s to 2.23s, and the BS offloading ratio increases from 46.37\% to 84.24\%. It can be seen from the simulation results that the proposed algorithm achieves better performance when the content popularity is more concentrated in the case of limited cache space of UTs.
\section{Conclusion}
In this paper, we have investigated the UT edge caching in D2D-enabled caching cellular networks.
The multi-winner auction for caching placement has been formulated. Based on which, a near optimal MOAC algorithm and a low-complexity MRAC algorithm have been proposed. It was demonstrated that the proposed algorithms solving the problem of user selfishness through the natural social efficiency and individual rationality of the proposed auction model.
Based on the user's preferences and the ``cache conflict" restrict relationship of the UTs, the proposed algorithms have selected different independent UT sets to cache different contents, which effectively avoids the waste of cache space. The  feasible and effective of the proposed algorithms have been verified by the simulation results.
In our future work, we try to optimize the radio resource allocation and caching placement jointly in a dynamic network with the aid of deep learning method.
\vspace{-1em}
\section*{Appendix~A: Proof of Lemma~\ref{lemma:optimal2}} \label{Appendix:A}
\renewcommand{\theequation}{D.\arabic{equation}}
\setcounter{equation}{0}
\begin{proof}
We assume the sets ${\cal W}$ and ${\cal V}$, that if and only if $\chi _n^* \ne 0$, $n \in {{\cal W}^*}$. Similarly, if and only if $y_n^* \ne 0$ , $n \in {{\cal V}^*}$. For the binary variable ${\chi _n}$ , the constraint ${\kern 1pt} {\kern 1pt} {\chi _n} + {\chi _{n'}} \le 1$ can also be expressed as ${\chi _n}{\chi _{n'}} = 0 $. We define ${y_n} = \frac{{\sqrt {{\upsilon _n}} {\chi _n}}}{{\sqrt {\sum\nolimits_{k \in {{\cal W}^*}} {{\upsilon _k}} } }}$ and ${\left| {\bf{y}} \right|_2} = 1$ on the basis of ${\chi _n}$, and for any UT $n$, $n'$, if ${E_{n,n'}} = 1$, then ${y_n}{y_{n'}} = \frac{{\sqrt {{\upsilon _n}{\upsilon _{n'}}} }}{{\sum\nolimits_{k \in {{\cal W}^*}} {{\upsilon _k}} }}{\chi _n}{\chi _{n'}} = 0$. ${{\bf{y}}^*}$ that satisfies the above constraints can be considered as a reasonable solution. Since   ${\tilde U^*}$ is the maximum value and ${\left| {\bf{y}} \right|_2} = 1$, we know that,
\begin{align}\label{optimal3-1}
\scriptsize
{\tilde U^*} \ge {({{\bf{\mu }}^T}{\bf{y}})^2} = {(\frac{{\sum\nolimits_{n \in {{\cal W}^*}} {{\upsilon _n}} }}{{\sqrt {\sum\nolimits_{k \in {{\cal W}^*}} {{\upsilon _k}} } }})^2} = \sum\limits_{n \in {{\cal W}^*}} {{\upsilon _n}}  = {U^*}
\end{align}

At the same time, because ${{\bf{y}}^{\rm{*}}}$ is the optimal solution of problem~\eqref{optimal3}, we can express~\eqref{optimal3} equivalently as the following problem about the set ${\cal V}$,
\begin{align}\label{optimal3-2}
\begin{array}{l}
\mathop {\max }\limits_{{y_n},n \in {{\cal V}^*}} (\sum\limits_{n \in {{\cal V}^*}} {\sqrt {{\upsilon _n}} {y_n}{)^2},} \\
{\rm{s}}{\rm{.t}}{\rm{.}}{\kern 1pt} {\kern 1pt} {\kern 1pt} {\kern 1pt} {\kern 1pt} {\kern 1pt} \sum\limits_{n \in {{\cal V}^*}} {{{({y_n})}^2} = 1.}
\end{array}
\end{align}
According to the Cauchy-Schwartz inequality, $(\sum\limits_{n \in {{\cal V}^*}} {\sqrt {{\upsilon _n}} {y_n}{)^2}}  \le (\sum\limits_{n \in {{\cal V}^*}} {{\upsilon _n}} )[\sum\limits_{n \in {{\cal V}^*}} {{{({\upsilon _n})}^2}]}  = \sum\limits_{n \in {{\cal V}^*}} {{\upsilon _n}}$ is obtained, and when $y_n^* = c\sqrt {{\upsilon _n}}$ , the equation holds. Since if ${E_{n,n'}} = 1$ , $y_n^*y_{n'}^* = \frac{{\sqrt {{\upsilon _n}{\upsilon _{n'}}} }}{{\sum\nolimits_{k \in {{\cal W}^*}} {{\upsilon _k}} }}\chi _n^*\chi _{n'}^* = 0$ , we know that ${\cal V}$ meets the user's no cache conflict condition, so we can conclude that,
\begin{align}\label{optimal3-3}
{\tilde U^*} = \sum\limits_{n \in {{\cal V}^*}} {{v_n}}  \le {U^*}
\end{align}
Comparing~\eqref{optimal3-1} and~\eqref{optimal3-3}, we can see that Lemma 1 has been proved.
\end{proof}

\begin{spacing}{1}
\bibliographystyle{IEEEtran}

\end{spacing}

\end{document}